\documentclass[]{aastex631}

\begin{document}

%\title{B-class flare and Filament Eruption from AR12661}
%\title{{\bf Formation of a Sigmoid, Partial Filament Eruption, and B-class Solar Flare via Tether-Cutting Reconnection in AR12661}}
\title{Sigmoid Formation, Filament Destabilization, and Initiation of Weak Flare by Tether-Cutting Reconnection}%, and Finally, Partial Filament Eruption Removed the Sigmoid.}

\correspondingauthor{Pradeep Kayshap}
\email{virat.com@gmail.com}

\author[0009-0007-1911-797X]{B. Suresh Babu}
\affiliation{School of Advanced Sciences and Languages, VIT Bhopal University, Kothri-Kalan, Sehore, 466114, M.P, India}

\author[0000-0002-1509-3970]{Pradeep kayshap}
\affiliation{School of Advanced Sciences and Languages, VIT Bhopal University, Kothri-Kalan, Sehore, 466114, M.P, India}

\author{Ashok Kumar Baral}
%\email{bnd.app@iitbhu.ac.in}
\affiliation{School of Advanced Sciences and Languages, VIT Bhopal University, Kothri-Kalan, Sehore, 466114, M.P, India}

\author{Bhola N. Dwivedi}
%\email{bnd.app@iitbhu.ac.in}
%\affiliation{Dept. of Physics (Retd. Professor), Indian Institute of Technology (BHU), Varanasi-221005, India}
\affiliation{Retired from Indian Institute of Technology (BHU), Varanasi-221005, India}

%\author{F.X Timmes}
%\affiliation{Arizona State University}
%\affiliation{AAS Journals Associate Editor-in-Chief}



\begin{abstract}
We have studied a B-class solar flare and an associated filament eruption through multi-wavelength observations. The flare triggers at 16:24~UT on June 7$^{th}$, 2017 from an active region (AR) 12661, and it maximizes at 16:54~UT. The magnetic flux cancellation occurs near the polarity inversion line (PIL) preceding the flare, and ultraviolet (UV) brightenings occur in the pre-flare phase at the flux cancellation sites, suggesting the reconnection occurs in the lower atmosphere, initially. The S-shaped sigmoid forms through successive steps in corona, i.e., small-scale brightenings, helical/twisted field lines, bright patches, and finally, a developed sigmoid. It justifies that runaway reconnection within the coronal arcades forms the sigmoid within the filament. The differential emission-measure (DEM) analysis reveals the existence of the plasma at a temperature of more than 10 MK within the sigmoid. The initial magnetic reconnection reorganizes the field overlying the filament as per the tether-cutting model. Therefore, it enables the filament to rise slowly, and around~16:41~UT, the eruption phase of the filament begins. The filament eruption removes the overlying coronal field, including the sigmoid. During the eruption phase, we have found intersecting/crossing of coronal loops and jet-like structures far away from the sigmoid-filament system. In conclusion, all the observational findings (e.g., magnetic flux convergence, cancellation, UV brightenings, and spatial and temporal correlation between formation/evolution of the sigmoid and rise/eruption of the filament) suggest that the formation of a solar flare and the eruption of the filament are consistent with the tether-cutting model of solar eruptions. 
\end{abstract}



\section{Introduction} \label{sec:intro}

Solar flares are gigantic explosions of energy in the solar atmosphere, and they are well-studied phenomena of the solar atmosphere (\citealt[and references cited therein]{2017LRSP...14....2B}). With time, significant progress has been made, and as a result, the classical standard model of solar flares came into existence, i.e., the CSHKP model (\citealt{1964NASSP..50..451C, 1966Natur.211..695S, 1974SoPh...34..323H, 1976SoPh...50...85K}). Later on, the 3D extension of the standard flare model is given by \cite{2012A&A...543A.110A}. Despite such significant advances, various aspects of solar flares remain unanswered or poorly understood, which need further investigations (e.g., \citealt[and references cited therein]{2011LRSP....8....6S, 2011LRSP....8....5A, 2011SSRv..159...19F, 2013AdSpR..52.1561C}). As per the CSHKP model, the reconnection occurs in the solar corona, and high-energy particles precipitate in the lower solar atmosphere, forming the flare ribbons (see \citealt{2017LRSP...14....2B}). Usually, two types of ribbons form in the lower solar atmosphere, namely, two parallel ribbons and circular ribbons. Two parallel ribbons occur in the bipolar magnetic field configuration, and it results from reconnection among the surrounding arcades underneath the erupting filament or sigmoid (e.g., \citealt{2007ApJ...669.1372L, 2010ApJ...708..314A}). While the circular ribbons form in the more complex quadrupolar magnetic field configuration, i.e., fan-spine magnetic field configuration. In this case, the reconnection occurs at the top (i.e., null point) of the dome (i.e., fan-spine magnetic configuration), and high-energy particles precipitate around the dome in the lower solar atmosphere forming circular ribbons (e.g., \citealt{2009ApJ...700..559M, 2009ApJ...696L..70L,  2013ApJ...771L..30J, 2014ApJ...792...40V, 2015ApJ...806..171Y}).\\

Different activities exist in the pre-flare phase, known as precursors. Some important precursors are flux-emergence (e.g., \citealt{1977ApJ...216..123H, 1995JGR...100.3355F}), flux convergence and cancellations (e.g., \citealt{2018ApJ...869...78C, 2021ApJ...921L..33Y}), filament activation/eruption (e.g., \citealt{2005ApJ...630.1148S, 2019ApJ...880...84S}), sigmoids (\citealt{1992PASJ...44L..63T}), plasmoids (\citealt{2001EP&S...53..473S}), and transients (small-scale) brightenings (e.g., \citealt{2004JKAS...37...41M, 2017LRSP...14....2B}). The brightenings in the pre-flare phase are the result of magnetic reconnection (\citealt{2004JKAS...37...41M}). Such small-scale brightenings can trigger solar flares, and they are also responsible for the reorganization of the magnetic connectivities. In such a scenario, the reconnection within the sheared magnetic field of the corona can produce a sigmoid (e.g., \citealt{2007SoPh..243...63G}). Sigmoids are S or inverse-S shaped structures and generally observed in soft X-rays (e.g., \citealt{1992PASJ...44L..63T, 1992PASJ...44L.123S}) but, sometimes, they might appear in the ultraviolet (UV) and extreme ultraviolet (EUV; \citealt{1997ApJ...491L..55S, 2002ApJ...574.1021G}). The sigmoids are located above the filaments, and both structures (sigmoid and filaments) lie above the polarity inversion lines (PILs, \citealt{2002SoPh..207..111P}). Sigmoids tend to be more eruptive than non-sigmoid regions. In a statistical analysis, \cite{1999GeoRL..26..627C} showed that 84\% sigmoids were eruptive whereas, in the case of non-sigmoid regions, only 50\% were eruptive. Therefore, the sigmoid regions are being considered as the proxy for solar eruptions (e.g., \citealt{1996ApJ...464L.199R, 2011LRSP....8....1C}).\\

Various models explain the formation/eruption of the sigmoid in the solar corona as summarized by \citealt{2007SoPh..246..365G}, namely, (1) arcade model, (2) kink-unstable flux ropes, (3) current layer/sheet driven by flux-rope, and (4) vortex-driven models. In the arcade model, the self-amplifying magnetic reconnection within the sheared magnetic arcade forms the S (inverse-S) shaped sigmoid, and the sigmoid erupted later on as per the tether-cutting model (e.g., \citealt{1992LNP...399...69M, 2001ApJ...552..833M}). Next, in the case of the kink-instability flux-rope model, magnetic flux-ropes can undergo kink-instability in various physical conditions, and such kink-unstable flux ropes are frequently observed in solar eruptions (e.g., \citealt{2005ApJ...630L..97T, 2013ApJ...770L...3K, 2023ApJ...945..113M}). Although, before the eruptions, kink-unstable flux ropes appear as a sigmoid in the soft X-rays (e.g., \citealt{1996ApJ...464L.199R}). In the third model (i.e., current layer/sheet driven by flux rope), transient sigmoidal brightenings are either triggered by magnetohydrodynamic (MHD) instability or by the catastrophe of a flux rope, and finally produce a sigmoid in the current sheet (\citealt{1999A&A...351..707T, 2003ApJ...589L.105F,2004ApJ...617..600G}). Lastly, as per the vortex model, the photospheric vortices can be injected into the coronal flux rope; as a result, these coronal flux ropes become twisted ropes and appear as sigmoids (e.g.,\citealt{1992PASJ...44L..63T, 2003A&A...406.1043T}). See \citealt{2007SoPh..246..365G} for more details on these different models of sigmoid formation. \\

%Apart from these precursors, the occurrence of small-scale brightening before flare/filament eruption is a less widely reported precursor (e.g., 

Apart from the sigmoid, the filaments, which are cool structures in the solar atmosphere, are usually associated with solar flares. \cite{2004ApJ...614.1054J} showed that more than 95\% solar flares are associated with the active-region filaments. %Whereas, quiescent filament eruptions are associated with only 27\% solar flares. 
Nature of the filament eruption changes from event to event, for instance, full filament eruption (e.g., \citealt{2008ApJ...674..586S,2010SoPh..261..127C, 2013ApJ...771...65J}), partial filament eruption (e.g., \citealt{2009A&A...498..295T, 2014ApJ...787...11J}), and the failed filament eruption (\citealt{2009ApJ...696L..70L}). Particularly, the failed eruption occurs due to various reasons, namely, (1) the dominance of overlying closed loops (e.g., \citealt{2013ApJ...778...70C, 2016PASJ...68....7X}), (2) asymmetric background magnetic field (\citealt{2009ApJ...696L..70L}), (3) small momentum of filament material due to non-potential active-region and weak Lorentz force (e.g., \citealt{2018ApJ...858..121L}), torus-stability region (e.g., \citealt{2005ApJ...630L..97T}), and gravity force (\citealt{2021PASA...38...18F}). Irrespective of the type of the eruption, the filament eruption can destabilize the magnetic field substantially and can trigger the solar flare (e.g., \citealt{2005ApJ...630.1148S, 2019ApJ...880...84S}). On the other hand, the filament eruption follows in time with flare emission, i.e., filament eruption occurs after the flare (\citealt{2012A&A...539A..27R}).
\cite{2001ApJ...554..474Z} showed small-scale brightenings near the filament before the eruption,  and they are co-spatial with cancelling magnetic features, i.e., the occurrence of the magnetic reconnection (e.g., \citealt{2007A&A...472..967C, 2011ApJ...731L...3S}). Similarly, in another work, small-scale brightenings before the filament eruption occur due to low atmospheric reconnection, which rapidly changes the chromospheric magnetic connectivity (\citealt{2001ApJ...547L..85K}). Finally, we mention that the reconnection (small-scale brightenings) in the pre-flare phase is an important observational signature that might help in the understanding of the eruption of the filament and formation of solar flares.\\
 
In the present work, we study a B-class solar flare and an associated filament eruption. Section~\ref{sec:obs} describes the observations. The results from multi-wavelength observations are described in Section~\ref{sec:mw_obs}. Finally, the discussion and conclusions are described in the last Section.\\

\section{Observations} \label{sec:obs}

The multi-wavelength observations have been used to investigate a B-class solar flare and an associated filament eruption. The study includes imaging and magnetic field observations. The imaging observations are taken from Global Oscillation Network Group (GONG; \citealt{1996Sci...272.1284H}), Atmospheric Imaging Assembly (AIA; \citealt{lemen2012}) onboard the Solar Dynamic Observatory (SDO; \citealt{2012SoPh..275....3P}), and X-Ray Telescope (XRT; \citealt{2007SoPh..243...63G}) onboard Hinode. While, the magnetic field observations are taken from the Helioseismic and Magnetic Imager (HMI; \citealt{2012SoPh..275..207S}) onboard SDO. 
%In addition, we also used the soft X-ray flux from the X-ray sensor (XRS) onboard Geostationary Operational Environment Satellites (GOES). 
An active region (AR) 12661, located near the disk center, is the site of the flare/filament eruption on June 07, 2017, from 14:00$-$19:30~UT. \\

%We have utilized chromospheric H-$\alpha$ images from the NSO/GONG, a network of six ground observatories globally. 
The chromospheric H-$\alpha$ observations from the Cerro Tololo Inter-American Observatory (CTIO) and Mauna Loa Observatory (MLO) are used in the present work. These H-$\alpha$ observations have a spatial resolution of 2$"$ and a temporal cadence of 1 minute (\citealt{1996Sci...272.1284H}). AIA/SDO provides high-resolution full-disk images of the Sun from the photosphere to the corona using 11 different filters with a spatial resolution of 0.6$"$ and a cadence of 12~s (\citealt{lemen2012}). The study particularly utilizes AIA~304~{\AA}, 171~{\AA}, and AIA~94~{\AA} images, which capture the emission from the transition region, lower corona, and corona, respectively. XRT captures the emissions in the soft X-ray wavelengths using different filters, namely, Al-poly, C-poly, Ti-poly, Be-thin, Be-med, Al-med, Al-thick, and Be-thick (\citealt{2007SoPh..243...63G}). The observations from the Be-thin filter, which captures the emission from a plasma at a very high temperature (i.e., log T = $\approx$ 7.0 K), are used in the present work.\\

Lastly, the HMI provides different types of magnetic field observation at the solar photosphere (\citealt{2012SoPh..275..207S}), and we have used a line-of-sight (LOS) full-disk magnetogram. It is to be noted that the spectroscopic diagnosis of this event has been carried out by \cite{2025Ap&SS.370...13B}. The Interface Region Imaging Spectrograph (IRIS) provides high-resolution near-ultraviolet (NUV) and far-ultraviolet (FUV) spectra containing various important spectral lines, such as Mg~{\sc ii}~h and k, C~{\sc ii}, Si~{\sc iv}, O~{\sc iv}, and many more (\citealt{Bart_2014}). The three important spectral lines (i.e., Mg~{\sc ii} k~2796.35~{\AA}, Si~{\sc iv}~1402.77~{\AA}, and O~{\sc iv}~1401.16~{\AA}) are also used by \cite{2025Ap&SS.370...13B} to investigate spectral intensity, Doppler velocity, non-thermal velocity, and line asymmetries during this B-class solar flare and filament eruption. 

\section{Multiwavelength Observations of Flare/Filament Eruption} \label{sec:mw_obs}
\subsection{Overview of the Event}
\begin{figure}
    \centering
    \mbox{
        \includegraphics[trim=4.5cm 0.0cm 2.0cm 0.3cm, scale = 0.90]{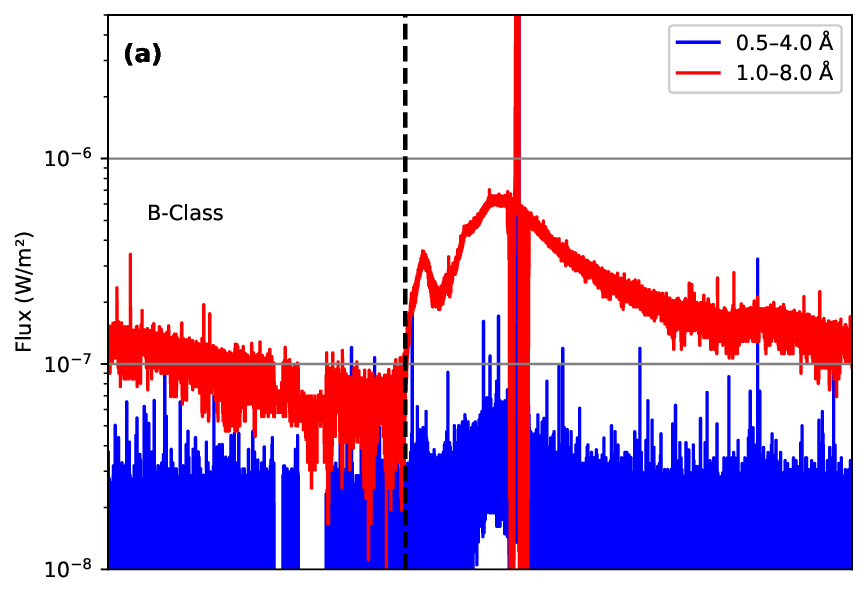}
        }
        \mbox{
        \includegraphics[trim=8.2cm 3.0cm 0.0cm 3.8cm, scale=1.75]{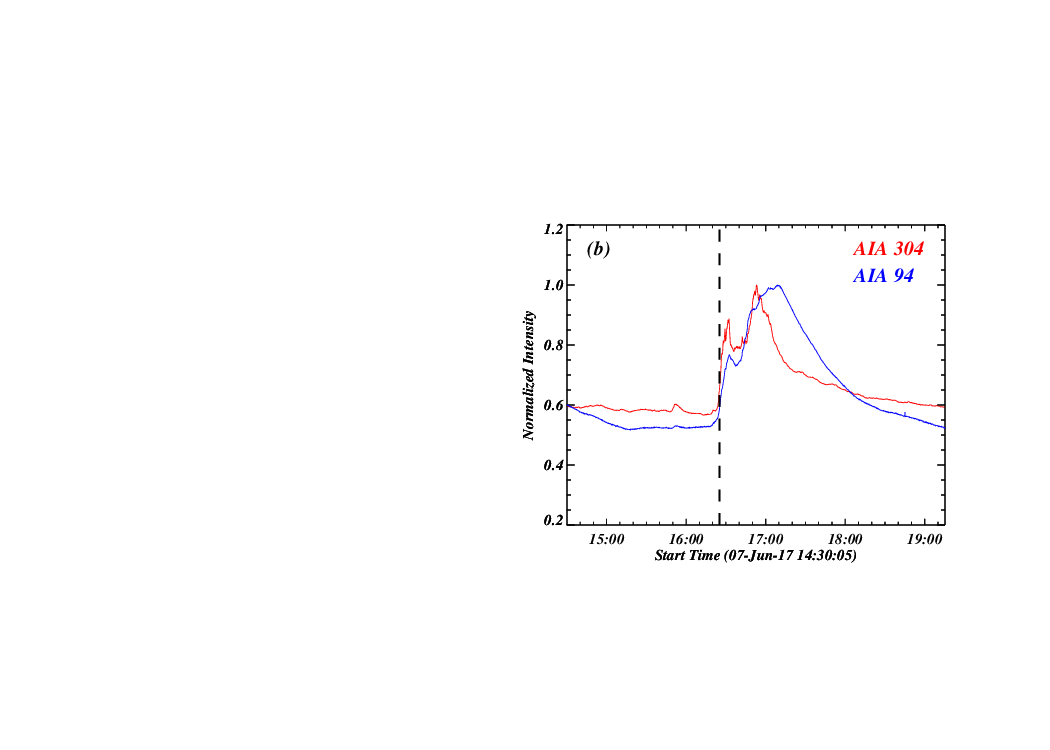}
        }
    \caption{The panel (a) shows the GOES soft X-ray (SXR; 1.0{-}8.0~{\AA}) flux profile (red curve) and hard X-ray (HXR; 0.5{-}4.0~{\AA}) flux profile (blue curve). The SXR flux profile starts to increase at 16:24~UT (vertical black dashed line), and peaks around 16:54~UT. The maximum flux is 7$\times$10$^{-7}$ W/m$^{2}$, therefore, it is a B-class solar flare. Panel (b) shows the flux profiles from AIA~304~{\AA} (red curve) and AIA~94~{\AA} (blue curve). The behaviour in these filters is consistent with the SXR flux profile.}
    \label{fig:euv_flux}
\end{figure}

The panel (a) of Figure~\ref{fig:euv_flux} shows the flux profiles from GOES soft X-ray (SXR; 1.0{--}8.0~{\AA}) and hard X-ray (HXR; 0.5{--}4.0~{\AA}) in red and blue colors, respectively.
%%%%%%%%%%%%%%% Fig: 1 %%%%%%%%%%%%%%%%%%%
%%%%%%%%%%%%%%%%%%%%%%%%%%%%%%%%%%%%%%%%%%%%%%%%%
The maximum of GOES SXR flux is around 7$\times$10$^{-7}$ W/m$^{2}$, and this maximum of SXR occurs around 16:54~UT, i.e., the flare gets its maximum phase around 16:54~UT. Around 19:30~UT, the SXR flux reaches the same level as it was before the triggering of the flare, i.e., the end of the solar flare. The panel (b) of figure~\ref{fig:euv_flux} shows the flux profile from two AIA filters, i.e., AIA~304~{\AA} (red profile) and AIA~94~{\AA} (blue profile). The behaviour of flare in AIA filters is consistent with the X-ray flux. The spectroscopic diagnosis of the same flare is done by \cite{2025Ap&SS.370...13B}, and they have also displayed the SXR and HXR of the same flare.\\

\begin{figure*}[hbt!]
\centering
\includegraphics[trim=2.5cm .0cm 3.0cm 0.5cm, scale=1.3]{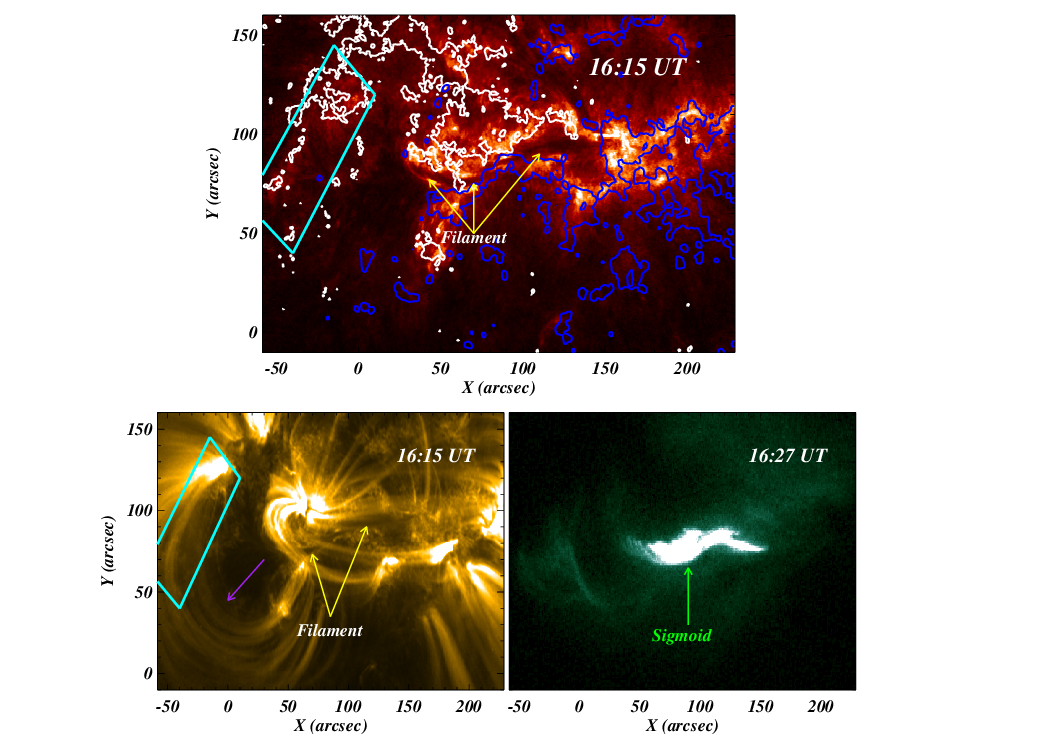}\\
\caption{The top panel shows AIA~304~{\AA} image from the pre-flare phase. The filament is indicated by yellow arrows, and it is located along the polarity inversion line. The overplotted contours are line-of-sight (LOS) magnetic fields obtained from HMI. The white (blue) contours show the positive (negative) polarity. A positive polarity region, located at a far-away location, is outlined by a cyan rectangular box. AIA~171~{\AA} image (bottom-left panel) shows that loops exist between this positive polarity and the negative polarity below the PIL. The filament is also visible in AIA~171~{\AA}, which is indicated by yellows in the bottom-left panel. The bottom-right panel displays the XRT/Be-thin image, which clearly shows the existence of a sigmoid.} 
\label{fig:ref_fig}
\end{figure*}

AIA~304~{\AA} image is displayed in the top panel of Figure~\ref{fig:ref_fig} from the pre-flare phase, i.e., at 16:15~UT. An elongated dark structure exists within the bright region which is indicated by yellow arrows. This dark elongated structure is the filament. Further, the contours on the AIA~304~{\AA} image show the line-of-sight (LOS) magnetic field from HMI. The white contours correspond to the positive polarity while the blue contours correspond to the negative polarity. It is visible that the filament lies along the boundary between positive and negative polarity, i.e., along the polarity inversion line (PIL). 
%%%%%%%%%%%%%%% Fig: 2 %%%%%%%%%%%%%%%%%%%
%%%%%%%%%%%%%%%%%%%%%%%%%%%%%%%%%%%%%%%%%%%
The cyan rectangular box on AIA~304~{\AA} outlines the positive polarity at a far location. The coronal loops exist between this positive polarity (inside the cyan rectangular box) and negative polarity below PIL, these loops are indicated by the purple arrow in the AIA~171~{\AA} images (bottom left panel). These loops participate in this event and remote brightening forms near the footpoint of these loops (see Section~\ref{sect:aia}). The same rectangular box is displayed in the bottom-left panel. The filament is also visible in AIA~171~{\AA} (indicated by yellow arrows), but weak compared to the AIA~304~{\AA}. Lastly, the bottom-right panel displays an image from the XRT/Be-thin filter after the solar flare. Here, the S-shaped sigmoid is developed and, like the filament, the sigmoid is also located along the PIL. Note that we have thoroughly checked the observations provided by Large Angle and Spectrometric Coronagraph (LASCO) onboard the Solar and Heliospheric Observatory (SOHO) to search for the coronal mass ejections (CMEs) associated with this event. We have not found any CMEs linked with this event.
%Figure \ref{fig:fig_goes}(b) shows the LOS magnetogram of AR12661 at 16:21~UT, i.e., prior to the flare. The black and white areas in the LOS magnetogram are negative and positive polarities, respectively. In the H-$\alpha$ image (panel c), the filament is surely visible, see the long dark structure indicated by cyan arrows. The filament location is extracted from the H-$\alpha$ image, and overplotted on the HMI magnetogram (green dashed line in panel b). Note that the green-dashed line lies on the border between positive and negative polarity, and this border is known as the polarity inversion line (PIL). Hence, the filament is located on the PIL, a common attribute of the filaments. In addition, the blue rectangular box in panel (b) outlines the positive polarity, away from the main magnetic field configuration. Note the magnetic loops exist from this positive polarity to the negative polarity below the PIL, and these magnetic loops participate in the dynamic of this event (see Section~\ref{sect:aia}).  

\subsection{Surface Magnetic Field}\label{sect:mag}
It is important to investigate the surface (photospheric) magnetic field to understand the triggering mechanism of this solar flare/filament eruption. 
%%%%%%%%%%%%%%% Fig: 10 %%%%%%%%%%%%%%%%%%%
\begin{figure*}[ht]
\centering
\includegraphics[trim=6.0cm -0.5cm 6.0cm 0.0cm, scale=1.10]{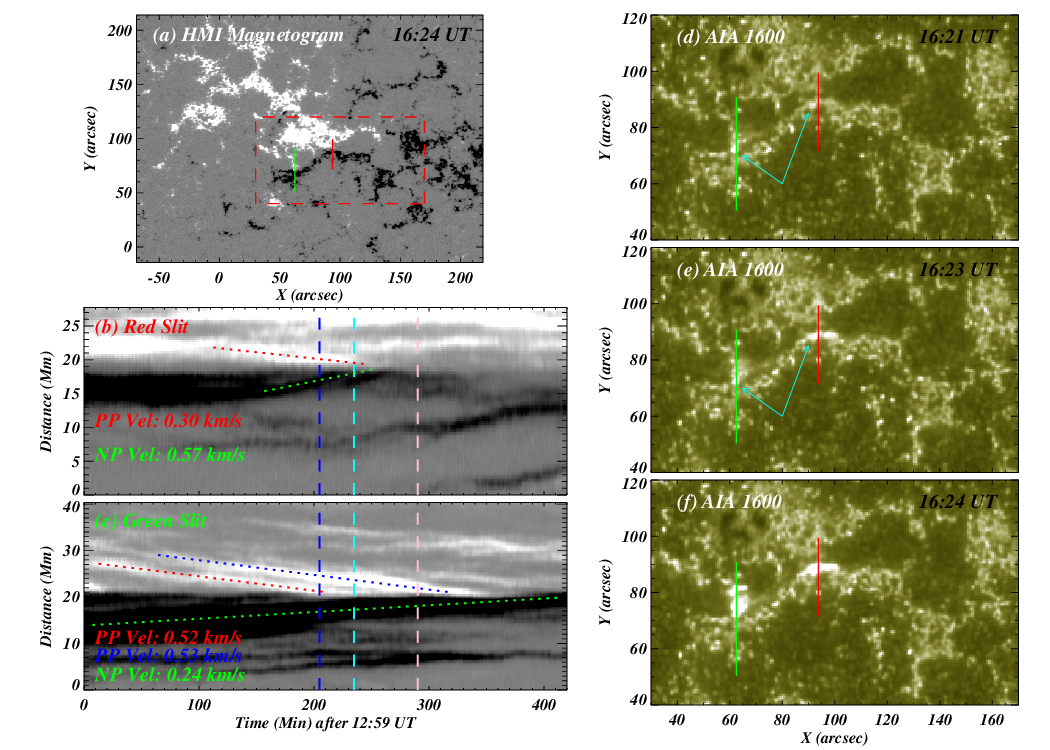}\\
\caption{The panel (a) shows the LOS magnetogram at 16:24~UT. The black (white) region corresponds to the negative (positive) polarity. The bigger red-dashed box outlines the flare region; the same box is displayed in the panel (e) of Figure~\ref{fig:fig_94}. The TD images produced from red and green slits (shown in panel (a)) are displayed in panels (b) and (c), respectively. The blue, cyan, and pink vertical dashed lines are located at the initiation, maximum, and end of the solar flares. In both TD images, it is explicitly visible that opposite polarities are approaching each other. Various paths are drawn on the positive and negative polarity patches (see slanted blue, red, and green dashed paths) to estimate their merging speeds. The panels (d), (e), and (f) show the intensity maps from AIA~1600~{\AA}. The compact brightnings are indicated by cyan arrows, and they are located under the red and green slits (same as shown in panel (a)).}
\label{fig:hmi}
\end{figure*}
%%%%%%%%%%%%%%%%%%%%%%%%%%%%%%%%%%%%%%%%%%%
We have used the HMI LOS magnetogram to investigate the surface magnetic field. The panel (a) of Figure~\ref{fig:hmi} shows the LOS magnetogram of the observed region, and the main flare region is outlined by the red dashed box.  It contains the positive (white) and negative (black) polarities. Here, we notice that the opposite polarities of the magnetic field are very close to each other in two different regions. We have placed two slits (see green and red slits in panel (a)) in these two regions to produce the time-distance (TD) images. The obtained TD images from red and green slits are displayed in panels (b) and (c), respectively. In both TD images, the blue, cyan, and pink dashed vertical lines are located at the times of initiation, maximum, and end of solar flares.\\

In panel (b), it is visible that both polarities are approaching each other. To estimate their convergence speed, we placed the paths along the positive (red-dashed line) and negative (green dashed path) polarities. The positive polarity is approaching the negative polarity at a speed of 0.30 km/s, and the negative polarity is approaching the positive polarity with a speed of 0.57 km/s. Please note that positive and negative polarities have met each other before the flare initiation (see both polarities before the blue line), and after this, positive flux gets weaker with time; it completely vanishes after the flare maximum (see the positive flux patch after the vertical cyan dashed line). This observational finding clearly shows the cancellation of magnetic flux, which starts before the flare and ends after the maximum of the flare.\\

The TD image from green slit (panel c) is more dynamics than the TD image from red slit (panel b). It shows a very strong negative polarity patch and multiple weak positive polarity patches. The negative polarity patch is converging towards positive polarity patches (see green dashed line) at a speed of 0.24 km/s. The convergence path of the first positive polarity patch is shown by the red dashed line, and the convergence speed is 0.52 km/s. Similar to the TD image deduced from the red-slit (panel b), the positive flux meets the negative flux before the time of the flare initiation, while it disappears after the maximum of the solar flare. There is another positive polarity patch that is converging towards the negative polarity, as outlined by the blue-dashed line. The convergence speed is very close to the first patch of positive polarity, i.e., 0.53 km/s. However, please note that this positive polarity patch interacts with negative polarity after the end of the flare; therefore, it might have produced activity after the flare as the positive flux patch disappears completely later on. Ultimately, this particular analysis shows flux cancellation in these two regions.\\

Not only surface magnetic field, but we have also examined the AIA~1600~{\AA} observations before the initiation of the solar flare (see panels (d), (e), and (f); Figure~\ref{fig:hmi}. The red and green slits, which are used to produce TD images from LOS magnetic field observation, are also shown in AIA~1600~{\AA} panels. The compact ultraviolet (UV) brightenings exist below these slits (indicated by cyan arrows), i.e., compact brightenings at the magnetic flux cancellation sites. Please note that these UV brightenings are expanding and getting stronger with time, see brightenings in panels (e) and (f). Actually, the magnetic cancellation is possibly the manifestation of the magnetic reconnection at the solar surface as reported by \cite{2018ApJ...861..135Y}, therefore, it means that the magnetic reconnection is getting stronger with time.

\subsection{Chromospheric observation}\label{sect:halpha}
\begin{figure*}[hbt!]
\centering
\includegraphics[trim=0cm 0.0cm 0.0cm 0.0cm, scale=1.0]{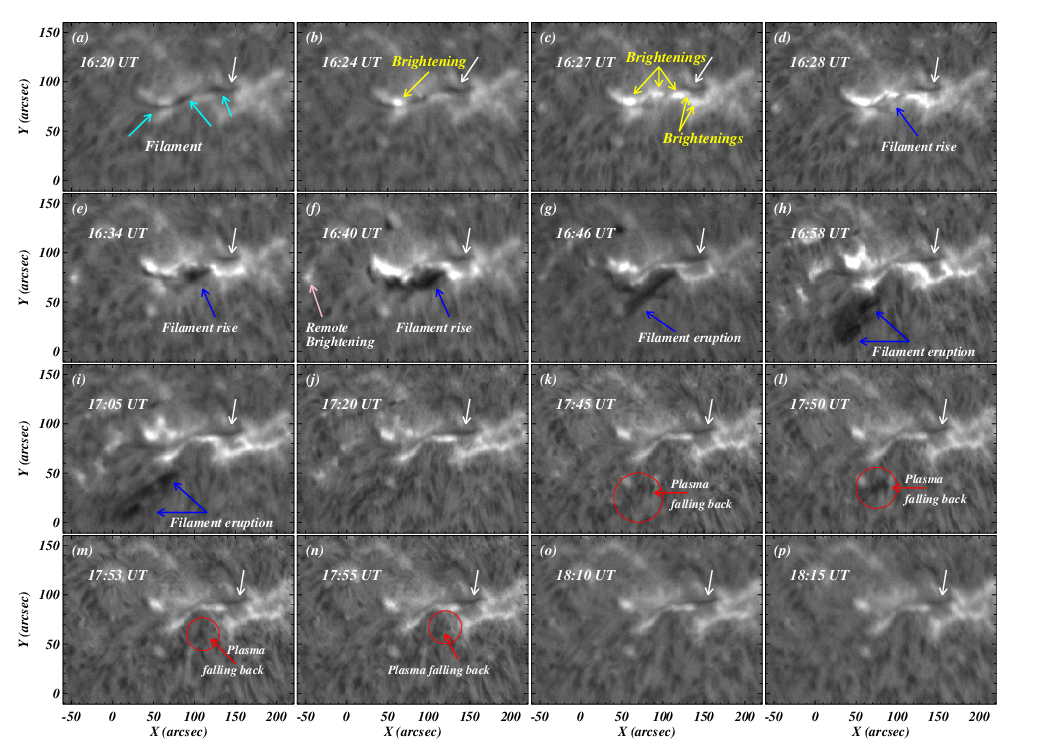}
\caption{The figure shows the evolution of the event using H-$\alpha$ observations. Several important features of the event are indicated by arrows, namely, filament (cyan arrows in panel (a)), brightenings around the filament (yellow arrows in panels (b) and (c)), filament rise (blue arrows in panels (d), (e), and (f)), filament eruption (blue arrows in (g), (h), and (i)) and plasma fall back (encircled by the red circle in panels (k) to (n)). The white arrow in each panel indicates the top part of the filament which does not change during the course of this event. The pink arrow in panel (f) indicates the remote brightening, which is located at the positive polarity location outlined by the cyan rectangular box in the top panel of Figure~\ref{fig:ref_fig}. Animation: The animation shows the temporal evolution of H-$\alpha$ and complements the figure shown here. The animation spans 16:15~UT to 18:15~UT at a cadence of 1 minute (20 frames per second; total duration $=$ 5 s). See animation halpha$\_$movie.mp4.}
\label{fig_halpha}
\end{figure*}

Figure~\ref{fig_halpha} shows the time evolution of the event in the H-$\alpha$ images. The cyan arrows indicate the filament at 16:20~UT (panel a). After around 04 minutes, a small but strong brightening exists near one edge of the filament, indicated by a yellow arrow in panel (b). This strong brightening at 16:24~UT is the chromospheric response of the solar flare, i.e., the formation of the H-$\alpha$ flare ribbon has started. 
%The GOES SXR flux shows the flare initiates at 16:24~UT (Figure~\ref{fig:fig_goes}(a)), and H-$\alpha$ observations also show the same time of flare initiation. 
Here, note that not only in H-$\alpha$ observations but we also see the triggering of a solar flare at the same time in AIA observations too (Section~\ref{sect:aia}). The solar flare is developing with time, so the brightenings in H-$\alpha$ images are also increasing. Therefore, after around 03 minutes (at 16:27~UT), five compact brightenings exist, which are indicated by yellow arrows in panel (c). Later, at 16:28~UT (panel d), the long bright patches exist, which are classical H-$alpha$ flare ribbons, and the H-$\alpha$ ribbons have covered almost the full filament at this time.\\

However, at 16:28~UT (panel d), a small dark structure is visible within the H-$\alpha$ ribbons indicated by the blue arrow. The flare hasn't reached its maximum till now, and therefore, the reduction in the size/brightness of the flare ribbons should not be expected. Therefore, the dark structure visible in panel (d) is the filament. It means the filament is rising now, so it appears above the flare ribbon. The filament rise (dark structure) is more clearly visible in the next panels (e) and (f), see the corresponding animation (halpha$\_$movie.mp4) for filament rise/eruption. Note that the filament rises phase lasts till $\approx$ 16:41~UT (see panels e and f), and then, the filament eruption phase around/after 16:41~UT. The filament eruption is clearly visible in the frames after 16:41~UT, see the dark elongated feature in panels (g), (h), and (i). At 17:05~UT, the filament material is not visible (panel i) because the filament material is out of the FOV. But after some time, some plasma remnants are falling back visible. These downfalling plasma remnants are circled and also indicated by the red arrows in panels (k), (l), (m), and (n).
%%%%%%%%%%%%%%% Fig: 3 %%%%%%%%%%%%%%%%%%%
%%%%%%%%%%%%%%%%%%%%%%%%%%%%%%%%%%%%%%%%%%%
It must be noted that some part of the filament, indicated by the white arrows, in all the panels, is unchanged during the whole event. Lastly, we mention a remote brightening, which is indicated by a pink arrow at 16:40~UT (panel f). This remote brightening is located at a site of positive polarity within a cyan rectangular box indicated in top and bottom-left panels of Figure~\ref{fig:ref_fig}(b).\\

%that the filament erupted partially because a \\

\begin{figure*}[ht]
\centering
\includegraphics[trim=4.0cm 4.0cm 4.0cm 3.0cm, scale=1.2]{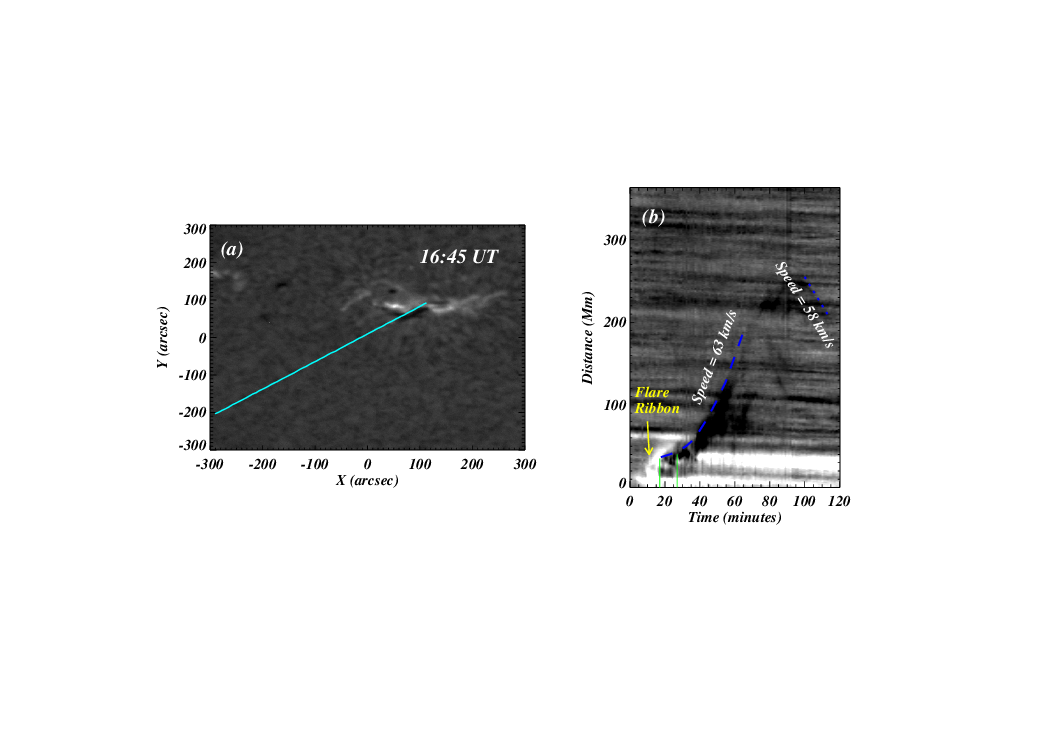}
\caption{Panel (a) shows the H-$\alpha$ image of the filament eruption and an artificial cyan slit is drawn along the path of the filament eruption. Then, the TD map is obtained using the cyan slit, which is shown in panel (b). The blue dotted curve in the TD map is drawn along the filament eruption. Firstly, the filament rises slowly (the slow-rising phase is enclosed within the vertical green lines), and later on, the filament erupts. The up-flow speed of the filament eruption (i.e., after the slow rise phase) is calculated, and it is 63 km/s. At the end of the event, the downfall of the plasma is visible, and a blue-dashed path is drawn along the downfall to estimate the downfall speed. The downfall speed is 58 km/s. The bright region before the filament rise motion is due to the flare ribbons.}
\label{fig:td}
\end{figure*}

The Time Distance (TD) analysis of filament eruption was performed using H-$\alpha$, which is shown in Figure \ref{fig:td}. We have drawn a path along the filament eruption (solid cyan line in panel (a)) to produce the TD image, and the produced TD image is displayed in panel (b). The upflow and the downflow of filament material is evident in the TD image. Further, we have drawn the path along the ascending motion of the filament, see the long dashed blue line in panel (b). Initially, the blue-dashed line rises slowly from 16:28~UT till 16:41~UT (region inside two vertical green lines), which shows the filament's rising phase. In this phase, the filament material rises only around 6 Mm (i.e., from 35 Mm to approximately 41 Mm) with a speed of $\approx$8 km/s. After 16:41~UT, the filament material ascends rapidly, i.e., the filament eruption phase. With the help of a long dashed blue line in panel (b) (after 16:42~UT), we measured the filament eruption speed which is 63 km/s. 
%%%%%%%%%%%%%%% Fig: 4 %%%%%%%%%%%%%%%%%%%
%%%%%%%%%%%%%%%%%%%%%%%%%%%%%%%%%%%%%%%%%%%
In the later phase, traces of the plasma downfall are visible, and a path is drawn along those traces in panel (b), see the dashed blue line in panel (b). The plasma downfall speed is around 58 km/s. Most importantly, the bright region (indicated by the yellow arrow in panel (b)) exists before the filament rise phase. This bright region is due to the presence of the H-$\alpha$ ribbon before filament rise/eruption.

\subsection{Transition-Region/Lower Coronal Observations}\label{sect:aia}
\begin{figure*}
\centering
\includegraphics[trim=0.0cm 0.5cm 0.0cm 1.0cm, scale=1.0]{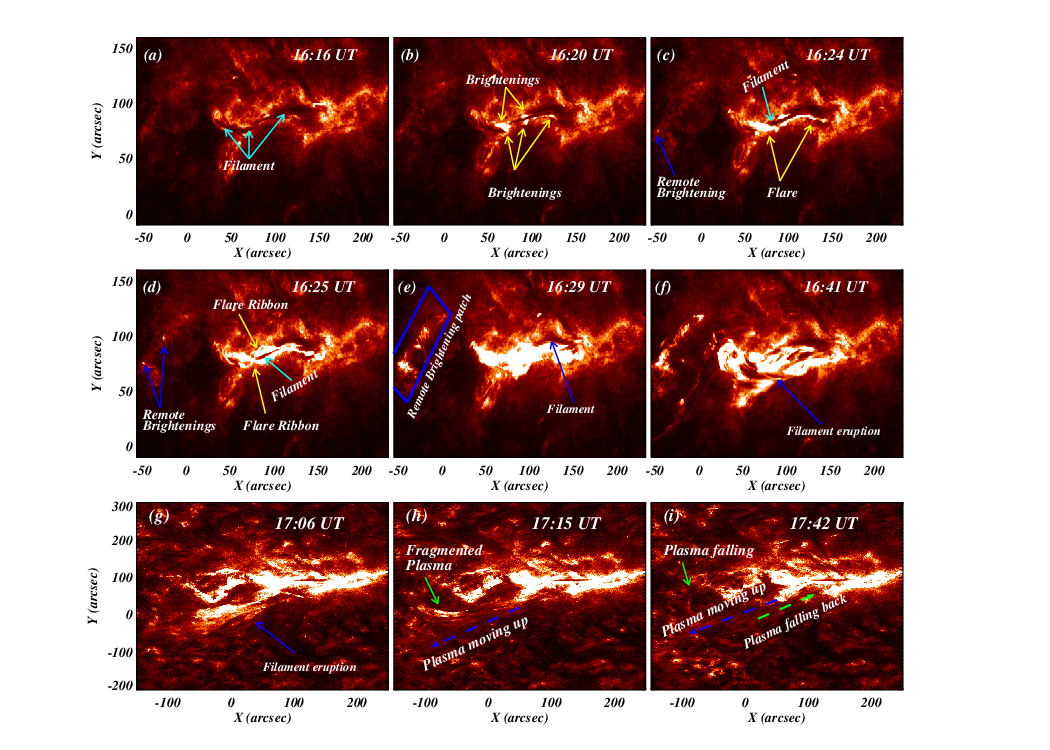}
\caption{The figure shows the evolution of event in the AIA~304~{\AA} channel. Various important observational features are found in this channel, e.g., filament (panel a), small-scale brightenings in the preflare phase (panel b), flare/flare ribbons (panels c and d), filament eruption (panels (f) and (g)), upflow, downflow and plasma fragmentation (panels g, h, and i), and remote brightenings (panels c, d, e, and f). Animation: The accompanying animation presents the temporal evolution of AIA 304~{\AA} emission over the same large field of view shown in the bottom row (panels (g)~{--}~(i)). The animation covers the eruption and post-eruption phases at a cadence of 12 s (20 frames per sec; total duration $=$ 5 s). The blue dashed arrow shows the upward motion of the filament eruption, the green arrow shows the fragmentation of plasma, and the yellow arrow shows the downward motion of the erupted filament plasma. see the animation aia304$\_$movie.mp4.}
\label{fig:fig_304}
\end{figure*}

\begin{figure*}
\centering
\includegraphics[trim=0.0cm 0.5cm 0.0cm 1.0cm, scale=1.0]{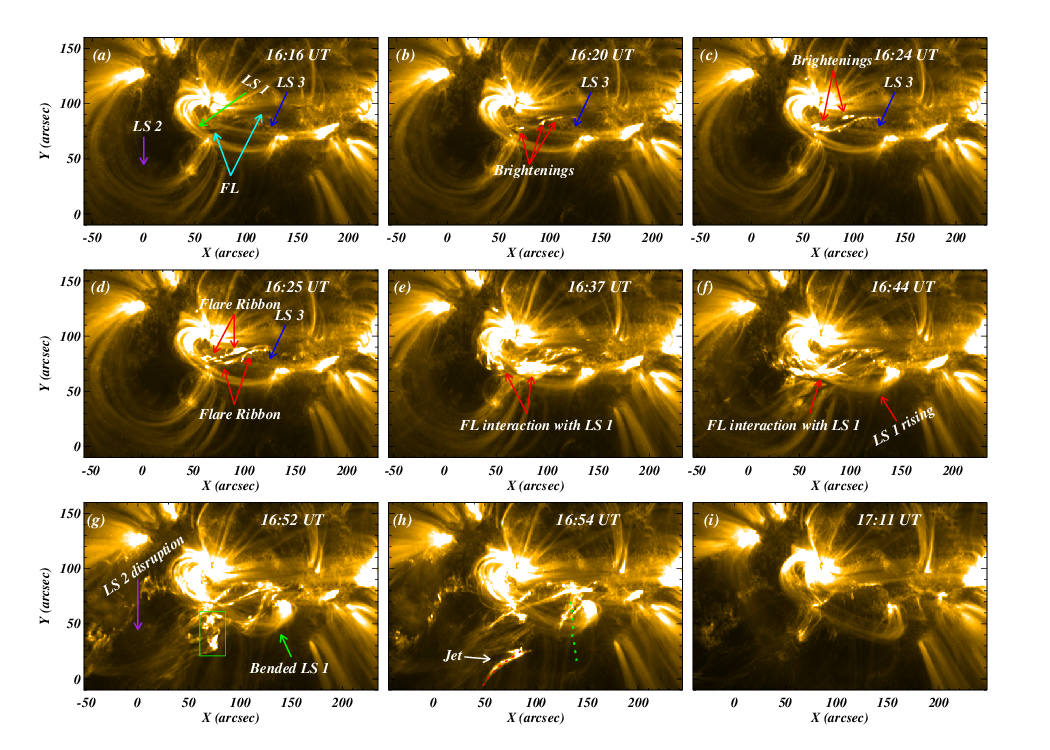}
\caption{Similar to AIA~304~{\AA}, filament (panels (a),(b), (c)), brightenings (panels (b) and (c)), and plasma upflows/downflows (panels (f), (g), and (h)) are also visible in AIA~171~{\AA}. Most importantly, we have identified three major loop systems (LS1, LS2, and LS3; panel (a)) that are affected by the flare/filament dynamics. Animation: The movie aia171$\_$movie.mp4 spans from 16:05~UT~{--}~19:15~UT covers the complete dynamics of the event at a cadence of 12 s (50 frames per sec; total duration $\approx$ 38 s).}
\label{fig:fig_171}
\end{figure*}

Figure~\ref{fig:fig_304} and~\ref{fig:fig_171} shows the evolution of the event in AIA~304~{\AA} and AIA~171~{\AA}, respectively. Before the solar flare, the filament is visible in AIA~304 (indicated by the cyan arrows in panel (a)). Firstly, the brightening at 16:18~UT exists (see movie aia304$\_$movie.mp4). In the next two minutes, more brightenings have developed around the filament (indicated by yellow arrows in panels (b)). The small-scale brightenings are distributed near the edges of filaments, and they continue to develop until the flare triggers at 16:24~UT (panel c). Just after one minute, two parallel flare ribbons have formed which are indicated by yellow arrows (panel d). The flare is well-developed at 16:29~UT (panel e). It must be noted that we didn't find any brightening in the H-$\alpha$ before the flare initiation, i.e., before 16:24~UT. The strong brightening at 16:24~UT in H-$\alpha$ Figure~\ref{fig_halpha}b is formed due to the flare initiation. Hence, we mention that the small-scale brightenings near the edges of the filament before the flare initiation are visible in the EUV wavelengths.
(i.e., AIA~304~{\AA} (Figure~\ref{fig:fig_304}) and 171~{\AA} observations (Figure~\ref{fig:fig_171})).\\
%are the result of the small-scale reconnections in the pre-flare phase.\\

At the flare-triggering time, we noticed a small-scale remote brightening far away from the flare/filament location; indicated by blue arrows in panels (c) and (d). This remote brightening continues to evolve with time and after around 5 minutes, it has become a large brightened patch, see the region within the blue rectangular box in the panel (e). This remote bright patch in AIA~304~{\AA} is at the same position as the position of the positive polarity patch in the HMI magnetogram which is outlined by the cyan rectangular box in the top panel of Figure~\ref{fig:ref_fig}. At t = 16:29~UT, the solar flare covered more than half of the filament. Only the top part of the filament is visible as indicated by the blue arrow in panel (e). The part of the filament covered by flare at 16:29~UT was visible earlier, see the filament indicated by a cyan arrow in panels (c) and (d). Later, the dark thread exists in the bright region at 16:41~UT indicated by the blue arrow in panel (f). The dark thread (i.e., cool plasma) is the erupted filament material. The filament eruption started around this time only (see Section~\ref{sect:halpha}). The solar flare attains the maximum phase around 16:54~UT, and it disappears completely around 19:30~UT.\\

The panels (g), (h), and (i) have bigger field-of-view than other panels of Figure~\ref{fig:fig_304}. The bigger FOV is chosen to show the complete path of filament material. The filament material has reached around x$\approx$ $-$100$"$ and y$\approx$ $-$50$"$ at t = 17:06~UT (panel g). Later on, the filament material is fragmented into two parts. One part follows a curved path (green arrow in panel (i)) and falls to a far location. While the other part continues to flow in the upward direction. The long blue-dashed arrow in panel (i) outlines the path of outflowing filament material. Finally, around 17:30~UT, the filament material stops around x$\approx$ $-$140 and y$\approx$ $-$100 and then starts to fall towards the solar surface, therefore, at t = 17:42~UT, the simultaneous existence of plasma downfall (green path) and upflow (blue path) is present. Please note the upflow and downflow of filament material are more clearly visible in the movie aia304$\_$movie.mp4. Similar to H-$\alpha$ observations, the top part of the filament is always present from the initial to the final phase of the event.\\

%%%%%%%%%%%%%%% Fig: 5 %%%%%%%%%%%%%%%%%%%
%%%%%%%%%%%%%%%%%%%%%%%%%%%%%%%%%%%%%%%%%%%

Next, the evolution and dynamics of this event in AIA~171~{\AA} (see Figure~\ref{fig:fig_171}) are consistent with AIA~304~{\AA}. In this filter, we again see all important observational signatures as seen in AIA~304~{\AA}, namely, filament (cyan arrows; panel a), small-scale brightenings (red arrows; panels (b) and (c)), two parallel flare ribbons (red arrows; panels (d)), and filament eruption (blue arrow; panel (g)). 

AIA~171~{\AA} is a good channel to investigate the evolution and dynamics of the coronal loops, i.e., it might help to understand the evolution/dynamics of the magnetic field for a particular activity. In the present observation of AR12661, we have identified three different loop systems which are affected due to this flare/filament eruption; loop system 1 (LS1; indicated by green arrow in panel (a)), loop system 2 (LS2; indicated by purple arrow in panel (a)), and loop system 3 (LS3; indicated by blue arrow in panel (a)). It seems that LS1 and LS3 are lying above the filament, while LS2 is lying below the filament. It is interesting to note that, firstly, LS3, which is the weakest loop system and located just above the filament, erupts around 16:37~UT (panel (e)). At this time, it is visible that the rising filament interacts with the LS1. Here, we mention that LS1 is also rising as the filament is rising (see Section~\ref{sect:fl_jet}). Later, it is visible that LS3 is completely removed (panel f), and the erupted filament material is interacting with LS1 (indicated by the red arrow in the panel (f)). Here, it should be noted that LS2 is still the same but stretched slightly outwards.\\   
%%%%%%%%%%%%%%% Fig: 6 %%%%%%%%%%%%%%%%%%%
%%%%%%%%%%%%%%%%%%%%%%%%%%%%%%%%%%%%%%%%%%%
At 16:52~UT, the erupted material of the filament is clearly visible, and it is disrupting the LS2. Meanwhile, the LS1 exists, but it is bent slightly as indicated by the green arrow. Therefore, the LS1 interacts partially with the erupted filament material. At this time, a vertically elongated bright patch is present (outlined by a green box; panel (g)), and after 2 minutes, a jet-like structure is formed near the bottom of the vertically elongated bright patch (panel (h)). In the end, all the material, including LS2 is swept away completely (dark region in panel (i)), but LS1 exists. However, this LS1 also fades after some time (not shown here).  %Note that the coronal loops (indicated by green arrows in panels (a), (b), and (c)) overlay the filament. Apart from these loops, we see some large curvature coronal loops (indicated by the purple arrow in panel (a)) in the nearby area. Large curvature loops were present until 16:40~UT (panel f), but after 16:41~UT, they started to disappear and completely disappeared at 16:54~UT (panel g). As we know, filament eruption starts around 16:41~UT, and the filament eruption has removed these large curvature loops. Therefore, the loops disappeared at 16:54~UT. 
Lastly, we mention that the signatures of the plasma fallback exist in AIA~171~{\AA}, but it is slightly weaker than in AIA~304~{\AA} (see movie aia171$\_$movie.mp4). 
 
\subsubsection{Filament Eruption and Jet-like Structure}
\label{sect:fl_jet}
The formation of a jet-like structure during the filament eruption is mentioned above. We have also mentioned that LS1 rises as the filament rises. To demonstrate the rising motion of LS1, we have produced TD image using a slit shown by the green dashed line in panel (h) of Figure~\ref{fig:fig_171}. 
%%%%%%%%%%%%%%% Fig: 6 %%%%%%%%%%%%%%%%%%%
\begin{figure*}
\centering
\mbox{
\includegraphics[trim=0.0cm 1.0cm 0.0cm 0.0cm, scale=1.0]{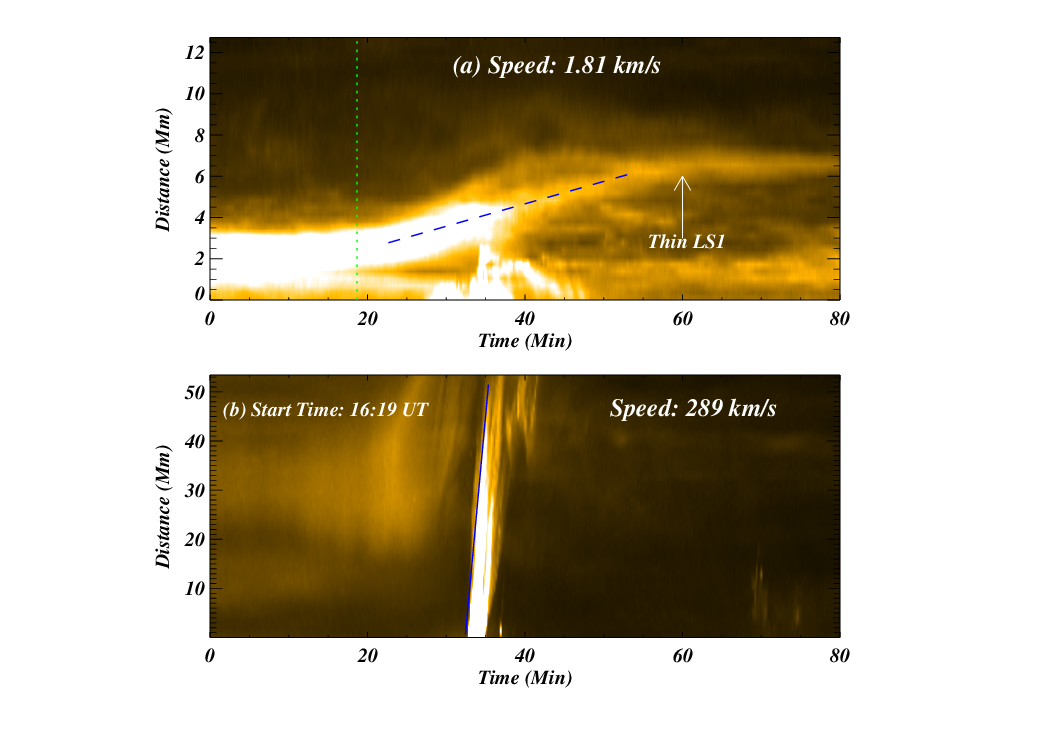}
}
\caption{The panels (a) and (b) display the time-distance images corresponding to red on the jet-like structure (panel (h); Figure~\ref{fig:fig_171}) and blue paths on the LS1 (panel (h); Figure~\ref{fig:fig_171}). The LS1 is rising with a speed of around 2 km/s (blue dashed path; panel (a)), and later on, the LS1 becomes a thin structure as indicated by the white arrow. The jet is moving with the speed of 289 km/s (blue solid line; panel (b)), and after the jet.}
\label{fig:td_jet}
\end{figure*}
%%%%%%%%%%%%%%%%%%%%%%%%%%%%%%%%%%%%%%%%%%%
The TD image is displayed in panel (a) of Figure~\ref{fig:td_jet}. Initially, the LS1 is rising slowly, i.e., see the region left to the vertical green dashed line. While around 16:38~UT, the LS1 starts to rise significantly. It is found that LS1 rises at a speed of around 2 km/s. Also, most importantly, the width of the LS1 is shrinking, which justifies that some of the loops (threads) of the LS1 have interacted with filament material. Lastly, LS1 appears as a very thin structure, as indicated by the white arrow.

This interaction of LS1 and erupted filament material produces an elongated bright patch (green box in panel (g); Figure~\ref{fig:fig_171}). Although not only the interaction between LS1 and erupted filament material, but it seems that interaction within the erupted filament material also contributes to the formation of the elongated bright patch. Hence, we mention that the vertical bright patch is the result of these two processes, namely, (1) interaction of erupted filament material with LS1 and (2) interaction within the erupted filament material. The jet is the result of these two different types of interactions, as the jet forms at the bottom of the elongated bright patch.\\

We have drawn the slit along the jet (see the red dashed line in panel (h); Figure~\ref{fig:fig_171}) to produce the TD image of the jet, and the corresponding TD image is displayed in the panel (b) of Figure~\ref{fig:td_jet}. The vertical bright strip around 35 minutes exists in the TD image, which is actually due to the upflow of the jet-like structure. A path is drawn (blue dashed line) along the ascending motion of the jet plasma, and with the help of this path, the speed of the jet is estimated. The speed of the jet is 289 km/s. The downfall of plasma is not visible because this jet-like structure is completely swept away with the erupted filament material. Not only this jet-like structure, but almost all the material from this region is swept away. Therefore, the emission in this region, after the jet-like structure, is significantly reduced; see the dark region after the bright strip in panel (b) of Figure~\ref{fig:td_jet}.  The jet-like structure has multi-thermal plasma, as it is visible in the various AIA filters (see appendix~\ref{append:jet}). Further, DEM of the base of the jet-like structure is displayed in Figure~\ref{fig:dem}, and it also confirms the existence of multi-thermal plasma at the base of the jet-like structure (see Section~\ref{sect:em}).

\subsection{Coronal Observations}\label{sect:xrt}
\begin{figure*}
\centering
\includegraphics[trim=0.0cm 1.5cm 0.0cm 2.0cm, scale=1.0]{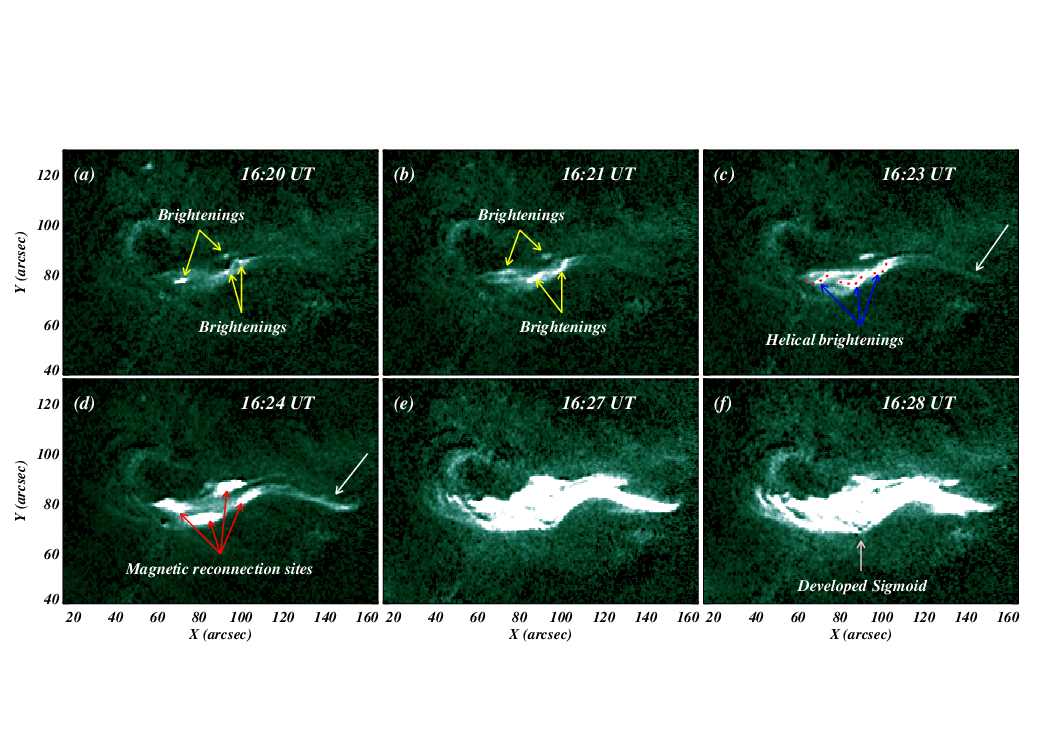}
\caption{Figure shows the formation of sigmoid using AIA~94~{\AA} observations. The small-scale brightenings are visible in the initial phase (indicated by yellow arrows in panels (a) and (b)). These brightenings transform into helical patches as outlined by red-dashed paths in panel (c). At this time, we noticed a curved leg indicated by the white arrow (panel c). Later, four well-developed brightenings exist, indicated by red arrows in panel (d). After 16:24~UT, instead of bright patches, the brightenings are distributed everywhere, and finally, the sigmoid appears (panels (e) and (f)).}
\label{fig:fig_94}
\end{figure*}

\begin{figure*}
\centering
\includegraphics[trim=3.0cm 0.1cm 3.0cm 0.2cm, scale=1.6]{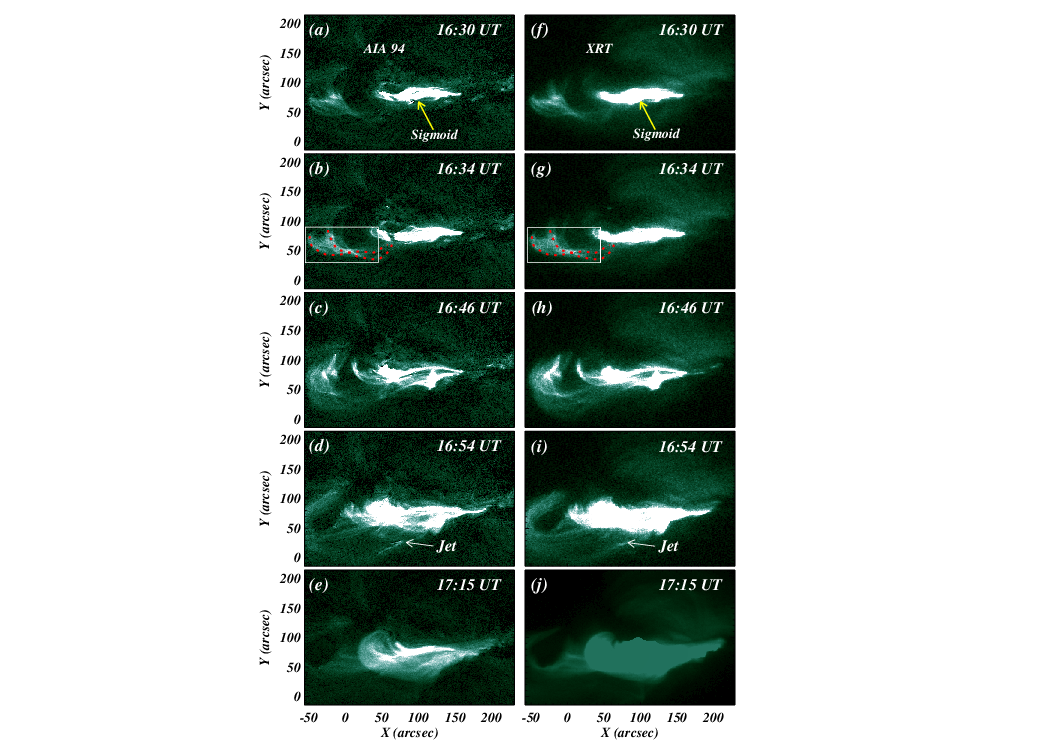}
\caption{The figure shows dynamics of the sigmoid, after its formation (Figure~\ref{fig:fig_94}), using AIA~94~{\AA} (panels (a) to (e)) and XRT/Be-thin (panels (f) to (j)). The sigmoid is visible in panels (a) and (b). As the event progressed, we noticed a brightening at a slightly far location (blue arrow panel (a)), which later transformed into a crossed bright loop structure, see red-dashed curve within the white rectangular box in panel (b). After the initiation of filament eruption (Section~\ref{sect:halpha}), the sigmoid is lost (panels (c), (d), and (e)), most probably, it is due to the interaction between filament and sigmoid as sigmoid lies above the filament. In addition, a compact bright region forms at 16:54~UT (blue box in panel (d)) due to filament interaction with the overlying field. The same dynamics are present in the XRT/Be-thin filter (panels (f) to (j)).}
\label{fig:fig_xrt}
\end{figure*}

The AIA~94~{\AA} is generally considered a hot filter as it captures the high-temperature plasma (i.e., log T/K 6.8; Fe~{\sc xviii} 93.932~{\AA})), but it can also capture the emission from relatively cool plasma (i.e., log T/K = 6.20; Fe~{\sc x} 94.012~{\AA})). To estimate the emission only due to Fe~{\sc xviii}, we used a method described by \cite{2013A&A...558A..73D} utilizing AIA~171 and 211~{\AA} as proxies along with AIA~94~{\AA} to remove the cool components (see the equation below).
%%%%%%%%%%%%%%%%%%%%%%%%%%%%%%%%%%%%%%%%%%%%%
\begin{center}
    %\begin{Math}
    \textit{I} (Fe~{\sc XVIII}) = \textit{I} (94~{\AA}) $-$ \textit{I} (211~{\AA})/120. $-$ \textit{I} (171~{\AA})/450.
%\end{Math}
\end{center}
%%%%%%%%%%%%%%%%%%%%%%%%%%%%%%%%%%%%%%%%%%%%%%%%
The AIA~94~{\AA} shows the sigmoid formation within the filament system (Figure~\ref{fig:fig_94}). At 16:20~UT, multiple bright dots exist (indicated by yellow arrows in panel (a)), and later, these bright dots become extended brightenings (panel (b)). Most importantly, at 16:23~UT, it is more clearly visible that these brightenings have developed along the curved path as outlined by red dotted lines (panel c). The development of brightenings along the curved path justifies that the inherited magnetic field lines are helical. Also, at this time, the extended leg has appeared as indicated by a white arrow, and the extended leg is more clearly visible at 16:25~UT (white arrow in panel d). Additionally, at 16:25~UT, four developed patches of brightening formed, indicated by red arrows (panel d). Later on, the brightening was enhanced and expanded in that region, and the S-shape sigmoid is fully developed  (panels (e) and (f)). Sigmoids are visible at high temperatures, and usually they are embedded/wrapping in the filament systems (e.g., \citealt{2018LRSP...15....7G, 2022ApJ...940...62L}).
%%%%%%%%%%%%%%% Fig: 7 %%%%%%%%%%%%%%%%%%%
%%%%%%%%%%%%%%%%%%%%%%%%%%%%%%%%%%%%%%%%%%%
Further, Figure~\ref{fig:fig_xrt} shows the dynamics of the sigmoid from 16:30~UT till 17:15~UT in the AIA~94~{\AA} (panels (a) to (e)) and XRT/Be-thin filters (panels (f) to (j)). The sigmoid, which is visible in panel (f) of Figure~\ref{fig:fig_94}, exists at 16:30~UT marked by the yellow arrow in Figure~\ref{fig:fig_xrt}(a). The sigmoid maintains its shape till 16:41~UT, but later, the sigmoid shape starts to distort. Please note that the filament eruption also starts after 16:41~UT. Further, the sigmoid is completely lost (panels (d) and  (e)).\\

%%%%%%%%%%%%%%% Fig: 8 %%%%%%%%%%%%%%%%%%%
%%%%%%%%%%%%%%%%%%%%%%%%%%%%%%%%%%%%%%%%%%%
Some loops are enhanced/formed at a slightly farther location from the main region, indicated by the blue arrow in panel (a). Later, at 16:34~UT, a long bright structure is present, and it looks like a typical X-type structure, see the red-dashed paths inside the white rectangular box in panel (b). During the rise phase, the coronal loops are lit up, and we believe that the groups of coronal loops are crossing (intersecting), which appears as an X-type structure. Hence, hereafter, this long bright structure is known as crossed (intersecting) bright loops. Please note that the crossed bright loops are visible till 16:46~UT. %Meanwhile, the far-location loops continue to brighten more and more (panels (a) and (b))}, and finally, we see large curvature loops as indicated by the blue arrow in panel (c). These large curvature loops are the same as visible in AIA~171~{\AA} (indicated by the pink arrow in Figure~\ref{fig:fig_171}(a)). After 16:46~UT, filament material interacted with the overlying coronal loops and removed overlying loops (see panels (d) and (e)). 
Please note the jet-like structure, which is explained in Section~\ref{sect:fl_jet}, is also visible in AIA~94~{\AA} (panel (d)) and XRT (panel (i)). 

%Most probably, the small bright region results from the interaction between overlying loops and filament material. This compact brightening has existed for some time and then erupted with the filament material.\\   

Like AIA~94, the Be-thin filter of XRT captures the emission from a very high temperature of log T/K = 7.0 (\citealt{2007SoPh..243...63G}). The panels from (f) to (j) of Figure~\ref{fig:fig_xrt} show the dynamics of the sigmoid in the XRT/Be-thin filter. The evolution and dynamics of the sigmoid in XRT are the same as seen in AIA~94~{\AA} (panels (a) to (e)). In addition, similar to AIA~94~{\AA}, the crossed bright loop structure (i.e., crossing of coronal loops) and compact small bright structure are also present in the XRT observations. 
%%%%%%%%%%%%%%%% Table 1 %%%%%%%%%%%%%%%%%%%%%%%%%%%%
\begin{table}[hbt!]
\centering
\caption{Onset Time of Key Features}
\label{table_onset}
\setlength{\tabcolsep}{3pt} % Adjust column spacing
\renewcommand{\arraystretch}{1.5} % Adjust row height
\noindent\begin{tabular}{|c|c|c|}
\hline
Sr. No & Observational Feature & Onset Time \\ 
\hline
1 & Small-Scale Brightenings & 16:18~UT \\
\hline
2 & Filament & Prior to 16:18~UT\\
\hline
3 & Flare Initiation & 16:24~UT\\
\hline
4 & Sigmoid Formation & 16:24~UT\\
\hline 
5 & Parallel Flare Ribbons & 16:25~UT\\
\hline
6 & Filament Rise Phase & 16:28{--}16:41~UT\\
\hline
7 & Filament Eruption Begins & 16:41~UT\\
\hline
8 & Flare Maximum & 16:54~UT\\
\hline
9 & Coronal Dimming & 17:50~UT\\
\hline
\end{tabular}
\end{table}
%%%%%%%%%%%%%%%%%%%%%%%%%%%%%%%%%%%%%%%%%%%%%%%%%
Lastly, the onset times of key features are estimated based on the observations described above, and they are mentioned in Table~\ref{table_onset}.

\subsection{Spatial and Temporal Connection Between Sigmoid Evolution and the Filament Destabilisation} \label{sect:Sigmoid_Filament}
\begin{figure*}
\centering
\includegraphics[trim=0.0cm -0.5cm 0.0cm -1.0cm, scale=1.0]{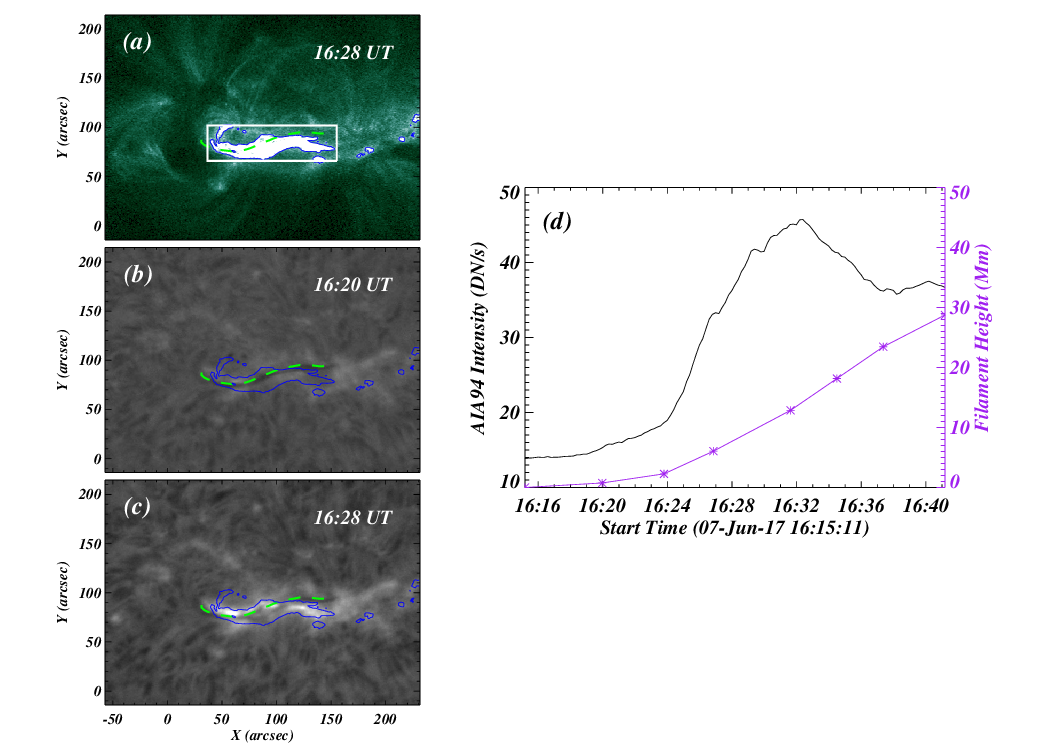}
\caption{The figure shows the spatial and temporal connection between the sigmoid and the filament/rise (i.e., filament destabilization). Panel (a) shows the fully developed sigmoid in AIA~94~{\AA} observed at 16:28~UT. Panel (b) shows the stable filament observed by H-$\alpha$ at 16:20~UT, and panel (c) shows the filament rise observed at 16:28~UT. A blue contour is drawn in a panel (a) to outline the sigmoid, and the same contour is drawn in panels (b) and (c). The initial position of the filament is tracked at time 16:20 (i.e., panel (b)), and displayed by the green dashed line. The same green dashed line is overplotted in the panels (b) and (c). This shows that the sigmoid is co-spatial with the filament. Panel (d) shows the temporal relation between the sigmoid evolution and the filament destabilisation (i.e., filament rise). The black curve in the panel (d) is the time evolution of AIA~94~{\AA} intensity (average intensity from the region within the white rectangular box in panel (a)) from 16:15~UT to 16:41~UT. The purple curve in panel (d) is the filament height traced for the same time duration. This timing analysis shows that the sigmoid evolution follows the filament destabilization.}
\label{fig:sigmoid_filamnet}
\end{figure*}

The panel (a) of Figure~\ref{fig:sigmoid_filamnet} shows the fully developed sigmoid observed by AIA~94~{\AA} at 16:28~UT. Next, the panels (b) and (c) show the H-$\alpha$ observation of a stable filament (observed at 16:20~UT) and the onset of filament rise (observed at 16:28~UT), respectively. In panel (a), we applied an intensity threshold (i.e., 30 DN/s) to draw the contour that outlines the sigmoid (see the blue contours). The same contour is drawn on the H-$\alpha$ observations, see blue contours in panels (b) and (c). Further, the green dashed line in panels (b) and (c) outlines the filament channel, and the same line is drawn on the sigmoid too (panel (a)). Now, it is clear that the sigmoid region covers a significant portion of the filament, see the green dashed lines within the blue contours in panels (b) and (c). The west leg of the filament is outside of the sigmoid region, and please note this part of the filament was unchanged during the whole event (section~\ref{sect:halpha}). Hence, it is confirmed that the sigmoid is co-spatial with the stable filament (before the eruption). Most probably, it provides the spatial linkage between the sigmoid and filament.\\
%%%%%%%%%%%%%%% Fig: 9 %%%%%%%%%%%%%%%%%%%
%%%%%%%%%%%%%%%%%%%%%%%%%%%%%%%%%%%%%%%%%%%%%%%%%%%%%%%%%%%%%%%
Next, the black curve in the panel (d) is the time evolution of the sigmoid emission (i.e., the averaged AIA~94~{\AA} intensity from the region within the white rectangle box in panel (a)) from 16:15~UT to 16:41~UT. Further, the height of the filament is traced for the same time duration (i.e., from 16:15~UT to 16:41~UT) and overplotted in panel (d), see the violet curve. This curve shows the filament rise phase before the eruption, i.e., destabilization of the filament. Initially, the AIA 94~{\AA} intensity begins to increase slowly at approximately 16:20~UT, marking the onset of sigmoid formation. The slow increase of AIA~94~{\AA} intensity lasts till 16:24~UT. Interestingly, the height of the filament changes from almost 0 Mm to nearly 2 Mm during these 4 minutes from 16:20 to 16:24~UT. After 16:24~UT, the sigmoid emission increases rapidly (i.e., the sigmoid forms rapidly), and the sigmoid is fully developed at 16:28~UT. And, as a result, the filament also rises rapidly after 16:24~UT, as the filament height changes from 2 Mm to 9 Mm in these 4 minutes. After the developed phase, the sigmoid starts to decay nearly at 16:32~UT, as the AIA~94 intensity is decreasing. However, the filament continues to rise, and the filament eruption starts at 16:41~UT. The panel(d) establishes the temporal linkage between the evolution of the sigmoid and filament destabilisation, i.e., formation/evolution of the sigmoid closely follows the filament destabilization.

%The spatial co-location of the hot sigmoid and filament, together with the clear temporal correlation between the evolution of the sigmoid and the destabilisation, suggests that the sigmoid is formed via a tether-cutting reconnection mechanism beneath the sheared filament, and leading to the slow rise and eventual partial eruption of the filament.}\\
\section{Emission-Measure Analysis} \label{sect:em}
\begin{figure*}

   \includegraphics[trim=0cm -0.10cm 0.0cm -1.0cm, scale=1.0]{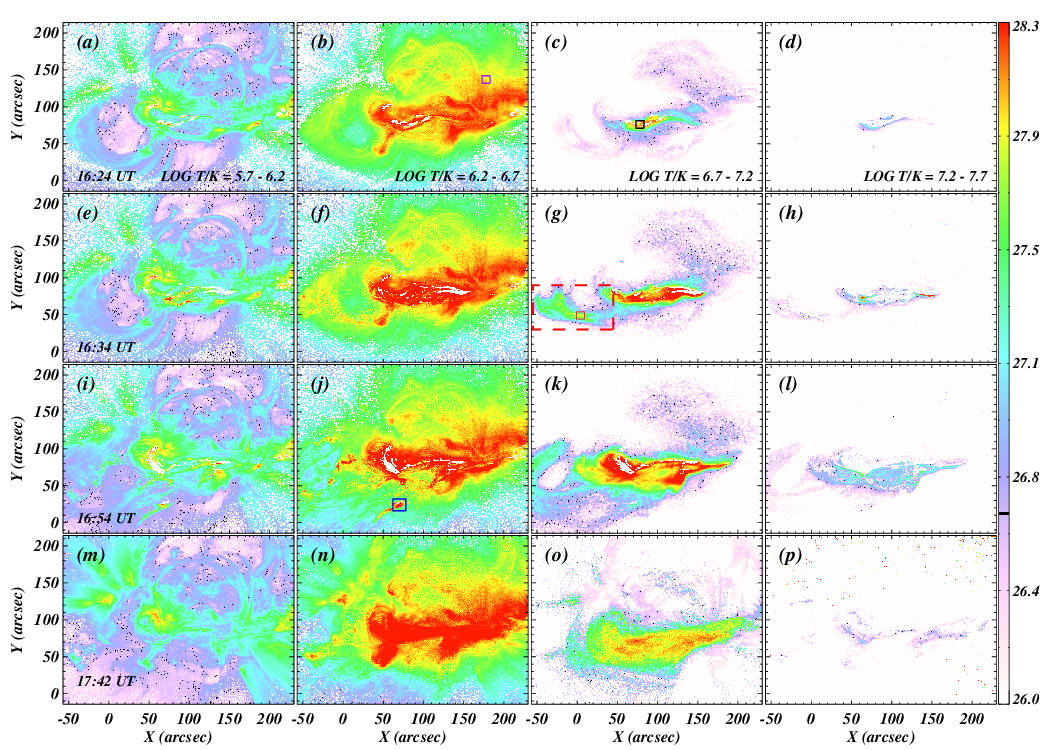}   
   \caption{The figure shows the EM maps obtained at four different times in four different temperature bins (T-bins), namely, (1) log T = 5.7{--}6.2 K, (2) log T = 6.2{--}6.7 K, (3) log T = 6.7{--}7.2 K, and (4) log T = 7.2{--}7.7 K. The emissions in the first two T-bins are significant at all four times (see the first two columns). The sigmoid has dominant emission in the third T-bin (see third column) and relatively faint emission in the last T-bin (see fourth column). The extended bright structure and compact bright region (see Section~\ref{sect:aia}) are mentioned here, see red and blue rectangular boxes in panels (g) and (k). In addition, we have selected small areas within the sigmoid and near the boundary of ARs, see the black and purple boxes in panels (c) and (b), respectively.} 
    \label{fig:em}
\end{figure*}

The emission measure (EM) is estimated using AIA and XRT (Be-Thin) observations with the method described by \cite{2015ApJ...807..143C}, i.e., aiaxrt.pro routine is used to estimate EM. Figure~\ref{fig:em} shows EM maps in four temperature bins (i.e., log T = 5.7{--}6.2 K, 6.2{--}6.7 K, 6.7{--}7.2 K, 7.2{--}7.7 K) at four different times, namely, (1) flare initiation (i.e., 16:24~UT; panels (a) to (d)), (2) 10 minutes after flare initiation (i.e., 16:34~UT; panels (e) to (h)), (3) flare maximum (i.e., 16:54~UT; panels (i) to (l)), and (4) decay phase of the flare (i.e., 17:42~UT; panels (m) to (p)). 
%%%%%%%%%%%%%%% Fig: 10 %%%%%%%%%%%%%%%%%%%
%%%%%%%%%%%%%%%%%%%%%%%%%%%%%%%%%%%%%%%%%%%%%%%%
The EM analysis shows significant emissions at all four times in the first two temperature bins, see the first two columns of Figure~\ref{fig:em}. At the flare initiation, the sigmoid has strong emission in the high-temperature range of log T = 6.7{--}7.2 K, and the rest of the area of AR12661 has diffuse emissions in this temperature range (panel c). Further, at the even higher temperature range (i.e., log T = 7.2{--}7.7 K), only the sigmoid shows emission, and there is no emission in the other part for the AR12261 (panel d). After 10 minutes of flare initiation, the spatial extent of the sigmoid significantly increased, and the emission also increased compared to the previous time (panels (g) and (h)). Again, the other area of ARs shows diffused emission in the temperature range of log T = 6.7{--}7.2 K (panel g), and almost no emission in the temperature range of log T=7.2{--}7.7~K (panel h). During the flare maximum, the high-temperature bins have strong emissions that have spread over a large area (panels (k) and (l)). Here, please also note that the sigmoid shape is lost at the flare maximum. Lastly, in the decay phase of the flare, the diffuse emission exists in log T = 6.9{--}7.2 (panel (o)), and very weak emission in log T = 7.2{--}7.7 (panel (p)) is sparsely distributed.\\

The emission is also enhanced at a far-away region, and the region is outlined by a red dashed rectangular box in panel (g). This is the same region that appeared as a crossed bright loop structure in the X-ray images (Section~\ref{sect:xrt}). Like X-ray images, the crossed/intersecting loops are also visible in the EM map of the temperature range log T = 6.7{--}7.2 K, see region inside the red dashed box in panel (g) of Figure~\ref{fig:em}. This particular structure has faint emission in the higher temperature bin, i.e.,  log T = 7.2{--}7.7 K. In addition, the jet-like structure is also visible in the panel (j).\\

%which is outlined by a {\bf blue box in panel (j)}. Lastly, we selected a small region within the sigmoid (black box; panel (c)) and small-scale brightenings (blue box; panel (k)). In addition, a small region is selected near the boundary of AR (purple box in panel c) for comparison.\\
%%%%%%%%%%DEM Analysis%%%%%%%%%%%%%%%%%%%%%%%%%

\begin{figure*}
\includegraphics[trim=-0.2cm 0.0cm 0.0cm 0.0cm, scale=0.97]{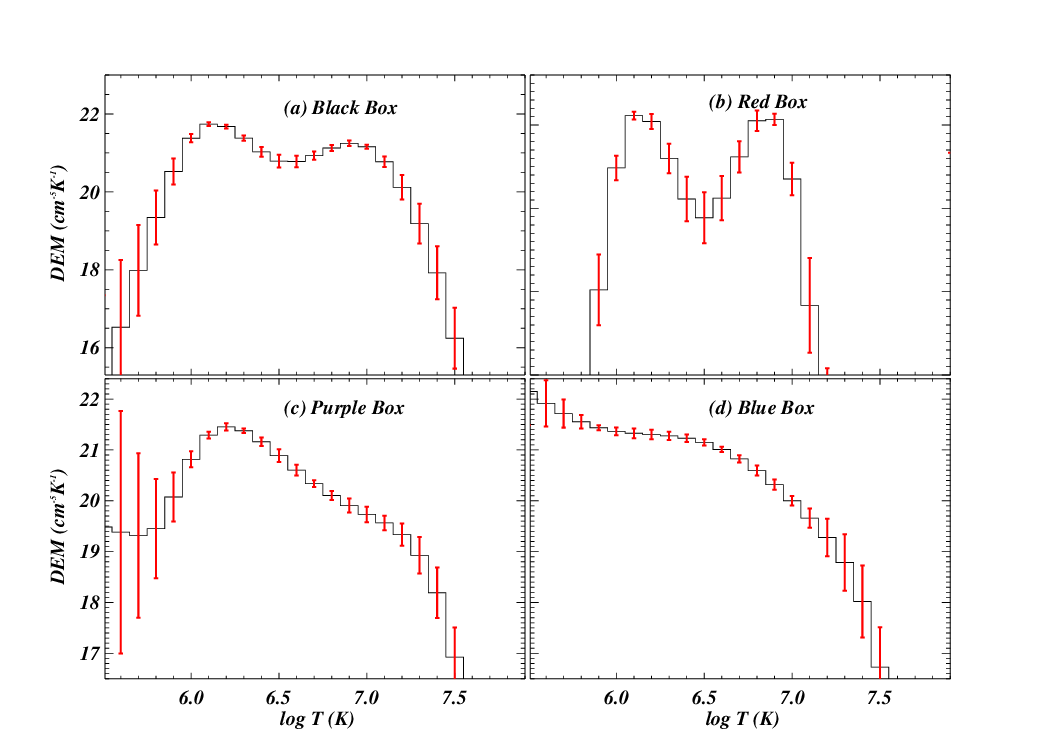}
\caption{The DEM from sigmoid (panel (a)), crossed loops (panel (b)), periphery of AR (panel (c)), and base of the jet-like structure (panel (d)). The vertical red bars are the standard deviation of DEM estimated from 100 Monte Carlo (MC) simulations.}   
\label{fig:dem}
\end{figure*}

The DEM estimation from joint AIA and XRT observations are more reliable towards the high temperature (\citealt{2021ApJ...915...96H}). To estimate the differential emission measure (DEM), we have used xrt$\_$dem$\_$iterative2.pro method (\citealt{2004IAUS..223..321W}) using intensities from AIA and XRT Be-Thin. This code is available in Solar Software (SSW). We have selected small regions within the sigmoid (black box in panel (c); Figure~\ref{fig:em}), crossed bright loops (small solid red box in panel (g); Figure~\ref{fig:em}), the periphery of ARs (purple rectangular box in panel (b); Figure~\ref{fig:em}), and jet-like structure (blue box in panel (j) of Figure~\ref{fig:em}) to estimate the DEM. All the intensities in each box are averaged in each filter, and the averaged intensity of the particular box is used to estimate the corresponding DEM. The Figure~\ref{fig:dem} shows the DEM curve from sigmoid (panel (a)), crossed bright loops (panel (b)), the periphery of AR (panel (c)), and the jet-like structure (panel (d)). The error (uncertainties) for each temperature bin is calculated as follows. The observed intensities are modified by adding random noise from a Gaussian. The random noise is multiplied by the errors in intensities, which is taken as 10\% of the observed intensities. And, finally, we have performed 100 Monte Carlo (MC) simulations on the observed intensities. Now, at each temperature bin, we have 100 DEM values, and the standard deviation of these 100 DEM values is considered as the corresponding error in that particular temperature bin.\\

%%%%%%%%%%%%%%% Fig: 9 %%%%%%%%%%%%%%%%%%%
%%%%%%%%%%%%%%%%%%%%%%%%%%%%%%%%%%%%%%%%%%%%%%%%
In the sigmoid region, the DEM peaks at log T = 6.1 K and log T = 7.0 K (panel (a)). The DEM falls sharply below log T = 5.7 K and above log T = 7.4 K. The error bars are small within the temperature range from log T = 5.7 K and log T = 7.4 K, while the error bars are very high at other temperatures. Similar to DEM from sigmoid, the DEM from crossed bright loops also has two separate peaks (panel (b)). The first peak is at the same temperature as seen in the sigmoid (i.e., log T = 6.1 K), while the second peak is at a relatively low temperature of log T = 6.8 K compared to the sigmoid.\\

On the contrary, the DEM from the periphery of AR is significantly different from the previous DEMs. The DEM peaks around log T = 6.2 K, similar to the first peak in the sigmoid and the crossed bright loops region. But after it, the DEM falls slowly till log T = 7.0 K, and further it falls rapidly. The error bars are large above log T = 7.0 and below log T = 5.8 K. The evolution and dynamics of the jet-like structure are described in Section~\ref{sect:fl_jet}. Now, we discuss the DEM (panel (d); Figure~\ref{fig:dem}) from the base region of this jet-like structure (blue box in panel (j); Figure~\ref{fig:em}). The clear peaks do not exist at the base of the jet-like structure, as seen in the DEMs from sigmoid and crossed bright loops. Rather, the DEM is falling slowly till log T = 7.0 K, and the DEM falls rapidly after log T = 7.0 K. Therefore, the jet-like structure has multithermal plasma, but the maximum emission of it is more in the low temperature.  

%%%%%%%%%%%%%%%%%%%%%%%%%%%%%%%%%%%%%%%%%%%
\section{Discussion and Conclusions}\label{sect:sec_sum}
A multiwavelength imaging analysis of a weak (B-class) solar flare and an associated filament eruption has been performed. The main findings of the work are as follows.
\begin{itemize}
      \item The flux cancellation is found in the two different locations within the flaring region. The compact UV brightnings also exist at the magnetic field cancellation sites. These are important observational signatures that justify the occurrence of magnetic reconnection via the tether-cutting (TC) model.

    \item The small-scale brightenings appeared in the pre-flare phase in high-temperature filters (AIA~94~{\AA}) and Hinode/XRT Be-thin, which finally resulted in the S-shaped sigmoid. In addition, we have also witnessed the small-scale brightenings in AIA~171~{\AA} and 304~{\AA} in the pre-flare phase. 
    
    \item The flare initiates at 16:24~UT on June 7$^{th}$, 2017 from AR12661. It reaches the maximum at 16:54~UT and completely disappears around 19:30~UT. The flare occurs in the simple bipolar magnetic field region. Therefore, parallel flare ribbons form in AIA~304~{\AA} and 171~{\AA}. In addition, a filament was present prior to the solar flare and below the sigmoid. 

    \item After the initiation of the solar flare, the filament starts to rise, and finally, the eruption of the filament begins around 16:41~UT. Initially, the filament material flows upward with a speed of around 63 km/s, and later on, the filament material fragments into two parts. Some filament material follows a curved path and falls in a far location. While other material falls back nearly on the same region with a speed of $\approx$ 58 km/s. 

    \item It is found that the S-shaped sigmoid and filament are co-spatial. Further, the formation/evolution of the sigmoid closely follows the rise of the filament (see Section~\ref{sect:Sigmoid_Filament}). Later, the partial eruption phase of the filament destroyed the sigmoid completely, as most of the material was swept away from the site. 
    
    \item During the slow rise of filament, the crossed bright loops appeared at a far location from the sigmoid filament system (white boxes in Figure~\ref{fig:fig_xrt}(b) and Figure~\ref{fig:fig_xrt}(g)). We believe that this structure is probably due to the intersecting/overlapping of the group of coronal loops during the rise phase of the flare. The DEM reveals the existence of multi-thermal plasma (Figure~\ref{fig:dem}(b)) %, and it might be the result of a brightened loop after the solar flare.} 
    
    \item During the filament eruption, a jet-like structure is formed (indicated by white arrow in Figure~\ref{fig:fig_xrt}(d) and Figure~\ref{fig:fig_xrt}(i). The jet plasma propagates outwards at the speed of 289 km/s, and it erupts completely with the filament material. Again, the plasma within the base of the jet-like structure is multithermal, and reaches up to the temperature of 10 million K as revealed by DEM (Figure~\ref{fig:dem}(d)). Here, please note that the sigmoid plasma also reaches that high temperature. Hence, most probably, this jet-like structure is a result of magnetic reconnection between the erupting field of the filament and the overlying coronal field.
\end{itemize}

Various models have been proposed for the formation of the sigmoid, and these models are nicely summarised by \cite{2007SoPh..246..365G}. One important model from them is the arcade model, which is motivated by the TC model for eruptive events (\citealt{2001ApJ...552..833M}). As per the arcade model, the core of sheared magnetic arcades is loaded with runaway magnetic reconnection that transforms reconnected field lines into twisted flux ropes (e.g., \citealt{1992LNP...399...69M, 
2001ApJ...552..833M}). The flux rope has a strongly sheared core in the middle and curves out to the opposite sides at the ends; therefore, they appear as S or inverse-S shaped structures, i.e., sigmoid (\citealt{2007SoPh..246..365G}). The initial magnetic reconnections within the coronal arcade can appear as small-scale brightenings. Such transient brightenings are important observational signatures, and they lead to the formation/heating of the sigmoid (e.g., \citealt{2000ApJ...532..628S, 2002ApJ...574.1021G, 2009A&A...498..295T, 2020A&A...644A.137J,2021ApJ...909...91K, 2021MNRAS.504.1201M}).\\

The magnetic flux convergence and cancellation start before the initiation of the solar flare and last after the maximum of the solar flare. Please note that before the solar flare/eruption, the convergence and cancellation of the photospheric magnetic flux is an important signature which justifies the flare/eruption occurring as per the TC model (e.g., \citealt{2018ApJ...869...78C, 2021ApJ...921L..33Y}). The cancellation of the surface magnetic field confirms the occurrence of magnetic reconnection at the solar photosphere (\citealt{2018ApJ...861..135Y}). Most importantly, UV brightenings exist at the magnetic flux cancellation sites (see Section~\ref{sect:mag}). Hence, these UV brightenings firmly establish the occurrence of magnetic reconnection below the filament as reported by \cite{2018ApJ...869...78C}. In the flux cancellation, one polarity (positive) completely disappears while the other polarity (negative) gets weaker with time from flare initiation to the flare maximum. Additionally, the UV brightenings are also increasing with time (see Section~\ref{sect:mag}). These observational findings suggest that, most likely, magnetic reconnection is getting stronger with time.\\ %The filament starts to rise due to the initial TC reconnection.}\\
 
The small-scale brightenings (Figure~\ref{fig:fig_94}(a) and Figure~\ref{fig:fig_94}(b)) exist in the solar corona. This is most probably the result of initial magnetic reconnection within the shared arcade, and later, these brightenings have evolved into helical bright paths (Figure~\ref{fig:fig_94}(c)). The helical bright paths justify the existence of the helical/twisted magnetic field lines because, in the highly structured corona, the plasma flows only along the magnetic field lines, and cross-diffusion is forbidden. Here, it must be noted that the occurrence of helical/twisted field lines is an important observational signature as per the TC model. A similar type of helical paths, in the solar corona, reported by \citealt{2018ApJ...869...78C}. Further, these helical paths are developed into the large bright patches (Figure~\ref{fig:fig_94}(d)). The magnetic flux cancellation and successive enhancement in the coronal intensities justify that the magnetic reconnection is getting stronger within the coronal arcade. Or, in other words, the implosive reconnection has changed to explosive reconnection as per the TC model (\cite{2001ApJ...552..833M}).\\

Finally, the bright patches have converted into a bigger bright S-shaped patch (i.e., sigmoid; Figure~\ref{fig:fig_94}(e)), which is the result of the explosive reconnection. \cite{2018ApJ...869...78C} have also mentioned that TC reconnection can convert the shared arcade into a twisted flux rope, i.e., sigmoid. The DEM of the sigmoid at 16:24~UT reveals the existence of high temperature plasma (i.e., around 10 million K; Figure~\ref{fig:dem}). Please note that with time, four bright patches (shown in Figure~\ref{fig:fig_94} (d)) develop into the sigmoid, and the DEM analysis shows that these four patches also have very high temperature plasma (like sigmoid). Such a high-temperature plasma in the corona is the result of the magnetic reconnection, as confirmed in the evolution of the photospheric magnetic flux. Hence, ultimately, the four bright patches are indeed the developed magnetic reconnection sites.\\    

Unlike the hot-temperature filters (i.e., AIA~94~{\AA} and XRT/Be-thin), the sigmoid is not present in relatively cool-temperature filters, such as AIA~171~{\AA} and AIA~304~{\AA}, but they show small-scale brightenings. The magnetic reconnection occurring within the coronal arcades deposits the high-energy particles into the lower atmosphere along the magnetic field lines, and they form the small-scale brightenings visible in AIA~304~{\AA} and AIA~171~{\AA}. Hence, the brightenings form near the edges of the filament because the magnetic field lines are connected at the edges of the filament over the PILs. Ultimately, the small-scale brightenings have transformed into parallel flare ribbons (Section~\ref{sect:aia})).\\

%%%%%%%%%%%%%%%%%%%%%%%%%%%%%%%%%%%%%%%%%%%%%%%%%
%Please note that magnetic field convergence, cancellation, UV brightenings, and helical field lines before the eruption are important decisive observational signatures that support TC reconnection process in the precursor phase as suggested by \citealt{2021ApJ...921L..33Y}.} 
The initial TC reconnection, which forms the small-scale brightenings prior to the eruption, can set up favorable conditions that are helpful in erupting the core field later, according to the TC model of solar eruptions. It should be noted that filaments are an integral part of the TC model to explain solar eruptions (\cite{2001ApJ...552..833M}), and in the present observation, the filament exists. As mentioned above, TC reconnection converted the shared arcades into the sigmoid; most probably, this sigmoid exists within the filament system, as the filament and sigmoid are almost co-spatial.
%It is found that the filament is present below the sigmoid: a common attribute of the filament-sigmoid system (\cite{2002ApJ...574.1021G}). 
This TC reconnection can reorganize the magnetic field lying above the filament (e.g., \citealt{2012ApJ...760...31K, 2013ApJ...773..128T}), or, most possibly, this reconnection removes the arcade lying above the filament \cite{2018ApJ...869...78C}. Most interestingly, the timing analysis shows that the formation/evolution of the sigmoid closely follows the slow rise phase of the filament. Hence, both phenomena (i.e., sigmoid formation and filament rise) are the result of the TC reconnection. The filament's slow rise phase lasts for about 13 minutes, and the filament rises with a speed of around 8 km/s (Section~\ref{sect:halpha}). Generally, the filament rise is considered the initial phase of filament eruption. The slow rise of the filament is followed by the eruption of the filament at a much higher speed of 63 km/s (Section~\ref{sect:halpha}). Please note that this filament eruption removes almost all coronal magnetic environments, including the sigmoids, in that vicinity. As a result, the coronal dimming region forms in the area of the interaction between the filament and coronal region (see panel (b) of Figure~\ref{fig:td_jet} and panels (h) and (i) of Figure~\ref{fig:fig_171}). In some cases, the filament eruption can trigger the solar flare, whereas in other cases, the flare (reconnection) can also trigger the filament eruption (e.g., \cite{2012ApJ...760...31K}). In the present observation, the filament erupts after the solar flares. Hence, the reconnection (solar flare) enables the filament eruption. The thorough analysis of this event yields several important observational findings, including magnetic field convergence and cancellation, UV brightenings preceding the flare/eruption, magnetic reconnection, the prior existence of a solar filament, helical field lines in the pre-flare phase, a sigmoidal shape, and the filament's slow rise before its eruption. All these observational findings firmly establish the formation of flare and filament eruption as per TC model (e.g., \citealt{kahler1988filament, 2001ApJ...552..833M, 2006SoPh..235..147Y, 2007ApJ...668..533N, 2007A&A...472..967C, 2007ApJ...669.1359S, 2011ApJ...731L...3S, 2014ApJ...797L..15C, 2018ApJ...869...78C, 2021ApJ...921L..33Y}).\\

Most widely, it is accepted that solar jets are the result of the magnetic reconnection occurring between two opposite magnetic field flux systems (e.g., \citealt{Shibata1992, 1996PASJ...48..353Y, 2013ApJ...763...24K, 2015Natur.523..437S, 2013ApJ...770L...3K, 2015A&A...581A.131J, 2017ApJ...849...78K, 2018A&A...616A..99K, 2021MNRAS.505.5311K}), and either the flux-emergence (e.g., \citealt{1996PASJ...48..353Y, Shibata2007, 2013ApJ...771...20M}) or plasma-instability (e.g., \citealt{2009ApJ...691...61P, 2023ApJ...945..113M}) trigger the magnetic reconnection and ultimately results in jet-like structures. However, apart from these processes, the eruption of the small-scale filaments (i.e., mini-filaments) also triggers the solar jets (e.g., \citealt{2015Natur.523..437S, 2018ApJ...854..122W, 2024SoPh..299...88K}). Next, for the first time, \cite{2015MNRAS.451.1117F} reported that not only mini-filaments, but eruption of a large filament triggers the solar jet, and the same is further reported by \citealt{2025ApJ...995....1Y}. Notably, in the present work, the interaction of the erupted filament and overlying coronal arcade produces the jet-like structure (Figure~\ref{fig:fig_171}(h)), and the jet-like structure is propagating with the speed of 289 km/s (Figure~\ref{fig:td_jet}). Next, we mention that CMEs are usually associated with the sigmoid and filament eruptions, although this is not always the case. Note that 32.4$\%$ of on-disk flare events had sigmoidal structures, but they did not have CMEs (\citealt{2018ApJ...869...99K}). Similar to the sigmoid, the filaments are usually associated with CMEs, but not always. \cite{2004ApJ...614.1054J} have shown that only 56\% filament eruptions are associated with the CMEs. Further, \cite{2011MNRAS.414.2803Y} have also reported that around 53\% filament eruptions are associated with CMEs. We have investigated the observation from LASCO/SoHO to know the presence of the CMEs with this event, and we didn't find any associated CME.\\

Not only the TC model but the magnetic breakout model and ideal MHD instabilities, such as the kink/torus instabilities, are also possible mechanisms that trigger solar eruptions (\citealt{1999ApJ...510..485A, 2012ApJ...760...31K, 2024ApJ...977..259F}). In the magnetic breakout model, the fan-spine magnetic topology is the main requirement, i.e., the quadrupolar magnetic field configuration should be present at the solar surface. The magnetic reconnection occurs at the null point in fan-spine topology (e.g., \citealt{1998ApJ...502L.181A, 1999ApJ...510..485A,2012ApJ...760...81K, 2024SoPh..299...88K}), and then, the energy from the higher atmosphere is transported into lower layers they produce circular/semi-circular flare ribbons (e.g., \citealt{2021MNRAS.501.4703J}). In the present observations, the quadrupolar magnetic field configuration and circular/semi-circular flare ribbons do not exist. Therefore, we rule out the possibility of the occurrence of magnetic breakout reconnection in support of the solar flare and eruption of the filament.\\ 

The kink instability is generally associated with strong rotational motion and/or writhing of the flux-rope/filament axis, often producing a characteristic "inverse $\gamma$" morphology (\citealt{2005ApJ...630L..97T, 2016NatCo...711837X, 2019ApJ...877L..28Z}). Although in this event, the S-shaped sigmoid develops after the initial low-lying magnetic reconnection, and sometimes, the sigmoids can be the signature of kink instability. But, this S-shape further develops into an inverse-$\gamma$ shape due to strong twist, which is not the case in the present observation. Hence, most probably, the S-shaped sigmoid is the result of shear in this event, not due to twist. Hence, the observational signature indicates that this eruption might not be triggered due to the kink instability. In the torus instability scenario, the outward hoop force of a flux rope is stronger than the downward magnetic tension force, i.e., torus instability occurs if the overlying magnetic field becomes too weak, and as a result, the flux rope rises. Hence, most of the time, the initial rise in torus instability occurs without magnetic reconnections. But in the present observation, the brightenings exist before the filament rise, i.e., magnetic reconnections occur in the very initial phase, and then the filament rises. Next, the eruption triggered by torus instability is strongly associated with CME; however, no CME is associated with this event. Hence, these observational signatures rule out the possibility of the occurrence of torus instability in the present event.\\

Lastly, we mention that initial magnetic reconnection occurs in the chromospheric heights along the polarity inversion line, as the initial brightenings are visible in cool temperature filters (i.e., IRIS/SJI~2796~{\AA}) before their existence in the hot coronal filters (image is not shown here). It justifies the occurrence of low-lying reconnection, which is a strong observational signature to support the occurrence of TC magnetic reconnection over the other alternative mechanisms. In conclusion, initial gentle reconnection within the arcade leads to the formation of the sigmoid, which simultaneously triggers a weak B-class flare and filament rise. Later, the gentle reconnection becomes explosive, and that triggers the main phase of the event (i.e., decay of sigmoid, filament eruption, and maximum of flare). The whole event is best explained as per the TC model.\\

\begin{acknowledgments}
The authors gratefully acknowledge the anonymous referee for his/her valuable and constructive comments that substantially improved the presentation of the manuscript.
BSB and PK thank R. L. Moore for the valuable discussion on this manuscript. We would like to thank the SDO, GONG, and GOES observations. 
\end{acknowledgments}

%%%%%%%%%%%%%Appendix%%%%%%%%%%%%%%%%%%%%%%%%%%%
\appendix
\renewcommand\thefigure{\thesection.\arabic{figure}}  
\section{Multithermal Nature of Jet-like Structure}
\label{append:jet}
Figure~\ref{fig:fig_jet_all} shows the jet-like structure (described in Section~\ref{sect:fl_jet}) in some AIA filters, namely, AIA~304~{\AA}~(Panel (a)), AIA~171~{\AA}~(Panel (b)), AIA~211~{\AA}~(Panel (c)), AIA~94~{\AA}~(Panel (d)), AIA~335~{\AA}~(Panel(e)) and XRT Be-Thin (panel (f)). Here, it should be noted that the AIA~94~{\AA} image shows emission only from the hot plasma because the contribution of the cool component is removed (\citealt{2013A&A...558A..73D})). The jet-like structure is visible in all the panels as indicated by the green arrow in all the panels. This suggests that the jet-like structure has multi-thermal plasma. 
%The coronal loops (LS2) are visible in AIA~171~{\AA} channel. LS2 rises as it interacts with the filament eruption. The green dashed line in panel (b) shows the path used to plot the HT image of the loop rising. The red dashed line on the jet-like structure in panel (b) shows the path used to plot the TD image of the jet evolution. The detailed explanation of the jet formation was given in \ref{sect:fl_jet}. 
%%%%%%%%%%%%%%% Fig: AppendiX 1%%%%%%%%%%%%%%%%%%%
\setcounter{figure}{0}
\begin{figure*}
\centering
\includegraphics[trim=1.0cm 0.8cm 0.0cm 0.0cm, scale=1.3]{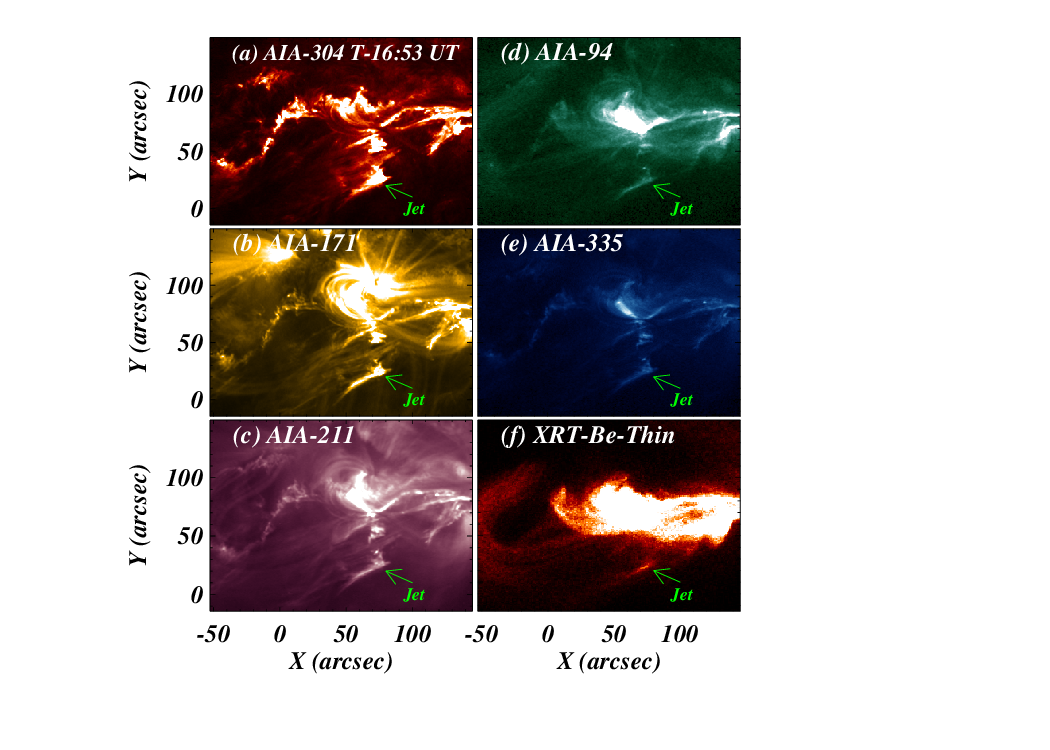}
\caption{The figure shows the jet-like structure in AIA~304~{\AA} (panel (a)), 171~{\AA}(panel (b)), 211~{\AA}(panel (c)), 94~{\AA}(panel (d)), 335~{\AA}(panel (e)), and XRT Be-Thin (Panel (f)). The green arrow indicates the jet-like structure in all the panels. Animation: The accompanying animation shows a zoomed-in view, focusing on the jet formation phase observed in AIA 171~{\AA}. The animation covers approximately 16:40~{--}~17:00~UT with a cadence of 12 s (displayed at 25 frames per sec; total duration $=$ 10 s). The cyan arrow initially shows the filament eruption. Later, the cyan arrow indicates plasma fragmentation at 16:50~UT. Finally, the jet formed at 16:54~UT (indicated by cyan arrow). The animation complements the static figure by isolating the jet evolution for better visualisation. see the animation aia171$\_$jet$\_$movie.mp4.}
\label{fig:fig_jet_all}
\end{figure*}
%%%%%%%%%%%%%%%%%%%%%%%%%%%%%%%%%%%%%%%%%%%

%\bibliography{references}{}
%\bibliographystyle{aasjournal}

{}



\end{document}